\documentclass[a4paper,11pt]{article}
  \usepackage[margin=1in]{geometry}
  \usepackage{amsmath,amstext,amssymb}
  \usepackage{graphicx}
  \usepackage[font=small]{caption}
  \usepackage[colorlinks]{hyperref}
  \usepackage[numbers,sort&compress]{natbib}
  \usepackage{doi} % For doi hyperlinks
  
  \usepackage[normalem]{ulem} % For \sout
  \usepackage{xcolor}
  
  \usepackage[T1]{fontenc} % Needed for \mathsfbi below
  \usepackage{lmodern} % Needed for T1 font encoding to work properly

\usepackage[utf8]{inputenc}
\usepackage{pict2e}
\usepackage{tikz}
\usetikzlibrary{decorations.pathreplacing}
\usepackage{svg}
\DeclareUnicodeCharacter{00D7}{$\times$} % Needed for svg

\newcommand{\I}{\mathrm{i}} % \sqrt{-1}
\newcommand{\e}{\mathrm{e}} % \exp\{1\}
\newcommand{\intd}{\mathrm{d}} % d/dx straight d
\newcommand{\vect}[1]{\boldsymbol{#1}} % Vectors
\DeclareMathAlphabet{\mathsfbi}{\encodingdefault}{\sfdefault}{bx}{sl} % Needed for mat def below
\newcommand{\mat}[1]{\mathsfbi{#1}} % Matrix
\DeclareMathOperator{\sgn}{sgn} % sgn sign function
\newcommand{\Real}{\mathrm{Re}}
\newcommand{\Imag}{\mathrm{Im}}

\newcommand{\wft}[4]{\widetilde{\ensuremath{#1}}^{(#2){#3}}_{{#4}}}%
\newcommand{\wdft}[4]{\widetilde{\ensuremath{#1}}^{(#2){#3}}_{{#4}}}%%

\begin{document}

  \title{\vspace{-0.5in}\textbf{Analysis of a rigid cylinder rolling over a linear elastic half-space in the full-slip regime}}
  \author{Hanson Bharth$^1$ and Edward James Brambley$^{1,2}$\footnote{Corresponding Author: \texttt{E.J.Brambley@warwick.ac.uk}}\\[1ex]%
  $^1$ Mathematics Institute, University of Warwick, UK\\%
  $^2$ WMG, University of Warwick, UK}

  \maketitle

\begin{abstract}
   This paper provides an analytical solution for the deformation of an elastic half-space caused by a cylindrical roller.  The roller is considered rigid, and is forced into the half space and rolls across its surface, with contact modelled by Coulomb friction.  In general, portions of the surface of the roller in contact with the half space may slip across the surface of the half space, or may stick to it.  In this paper, we consider only the regime where all of the rollers contact surface is slipping.  This results in a mixed boundary value problem, which is formulated as a $2\times 2$ matrix Wiener--Hopf problem.  The exponential factors in the Wiener--Hopf matrix allows a solution by following the iterative method of \citeauthor*{Priddin} (\emph{Phil.\ Trans.\ Roy.\ Soc.~A}~378, p.~20190241, 2020) which is implemented numerically by computing Cauchy transforms using a spectral method following \citeauthor*{slevinsky2017fast} (\emph{J.~Comput.\ Phys.}~332, pp.~290--315, 2017).  The limits of the contact region are located a posteriori by applying an optimisation method.  The solution is illustrated with several examples, and numerical code to compute the solutions in general is included in the supplementary material.
\end{abstract}

\section{Introduction}\label{sec:Intro}

The understanding of elastic deformation is a necessary basic step towards the further understanding and modelling of elastoplastic deformations, common in manufacturing processes.  As an example, strip rolling is a metal forming processes involving a pair of rollers squeezing and thinning a sheet of metal.  Provided the sheet is sufficiently wide, the process can be modelled as two-dimensional (plane strain) and steady state, and purely plastic models exist that agree well with finite element simulations~\citep{cawthorn+minton+brambley-2016,minton+cawthorn+brambley-2016}.  However, the neglect of elasticity precludes the modelling of important effects such as springback and curvature; indeed, the curvature of a sheet passing between two asymmetric rollers is currently an unsolved problem, with no agreement between experimental, computational and mathematical studies~\citep{minton+brambley-2017-ictp}.  This motivates the present study, investigating the vastly simplified situation consisting of a single roller rolling along a purely-elastic half-space, from which it is hoped understanding and further modelling can be developed that will contribute to future studies of elastoplastic deformation.

The model of an elastic half-space deformed by a rigid punch has been of interest for a considerable time.  \Citet{Hertz} founded the field of contact mechanics at the end of the 19th century by considering the elastic deformation due to a frictionless rigid punch.  Half a century later, the understanding of contact mechanics was extended when \citet{Cattaneo} and \citet{Mindlin} both considered elastic deformation of two cylinders under friction, setting the boundary conditions as an inner stick region and outer slip regions. \citet{Ciavarella1998} generalised Cattaneo's contact problem at the end of the 20th century, by formulating the problem in the slip regions as integral equations rather than conditions on displacement. 
The development of adhesive contact problems was led by \citet{MOSSAKOVSKII}, who used an incremental approach which was later used in both used in \citet{Goodman} and \citet{Borodich2004}. Alternatively, \citet{Spence} showed that a self-similar approach for contact problems may be used instead the incremental one, with Spence's approach making use of the Wiener--Hopf technique also. Spence formulated this approach by considering parabolic indenters and found the self-similar property to be the ratio of the stick region to the entire contact region remained the constant. The use of self-similarity has been implemented further in \citet{Borodich2002}, which investigated the effects of punch shape and frictional behaviour.  More recently, \citet{Zhupanska,Zhupanska_stick} modelled the deformation of an elastic half-space due to a rigid cylindrical indenter. However, due to the hysteric nature of friction, it is likely that the solution found in this case is not unique, as a cylinder pushed into an elastic half-space would produce a different deformation pattern from a cylinder pushed further into the elastic half-space before being partially retracted.
By considering here a rigid cylinder rolling across the surface of the elastic half space, the entire history of the deformation is specified within the model, and so a unique steady-state solution is to be expected.  In the limit of zero rolling speed, this solution should reduce to one solution to the non-rolling problem.

There are many possible models of friction between surfaces in contact.  Common models in metal forming include Coulomb friction, where the tangential force is proportional to the normal force, relative slip, where the tangential force is proportional to the slipping velocity, and ``friction factor'', where the tangential force is a specified constant.  The Coulomb friction model is the simplest friction model that is also well established outside of metal forming, and it is the friction model used here.  Under Coulomb friction, two surfaces in contact can be in one of two states: slipping, where the tangential force $T$ resists the slipping and is proportional to the normal force $N$, $T = \mu N$; and non-slipping, where $|T| < \mu N$ and the surfaces do not move with respect to one another.  In general, for a rigid cylinder rolling along and indenting an elastic half space, some of the contact surface will slip in one direction, some will slip in the other direction, and in between will be a region of no slip.  This general situation would result in a $4\times 4$ Wiener--Hopf problem owing to the four points where the boundary conditions on the elastic half space change.  Instead, we concentrate here on the regime where all of the contact surface slips in one direction.  This results in a $2\times 2$ Wiener--Hopf problem owing to the two points where the boundary conditions on the elastic half space change, as will be seen below, which is a considerable simplification.  This regime, which by analogy to a car may be thought of as a ``wheel spin'' or ``locked braking wheel'' regime, is largely ignored in the literature, although it is discussed by \citet{Sullivan} and \citet{Wang}.  It is hoped this $2\times 2$ regime, while mathematically interesting in its own right, will help with the development of a more complicated general $4\times 4$ model of stick-slip rolling.  It is worth noting that the conformal mapping method used by \citet{Zhupanska} results in a scalar ($1\times 1$) Wiener--Hopf problem, but is difficult to generalise to the rolling cylinder case studied here and does not generalise to the stick-slip case.

The problem posed here will turn out to result in a $2\times 2$ matrix Wiener--Hopf problem, amenable to solutions using various methods based on the Wiener--Hopf technique~\citep[e.g.][]{Noble}.
Such solution methods are well understood in the $1\times 1$ scalar case~\citep[e.g.][]{kisil2013constructive}, but are more difficult in the matrix case, and no universal method of solution is known; a review of approaches to solving matrix Wiener--Hopf problems is given by \citet{rogosin2016constructive}.  In particular, there is only a small class of matrix Wiener--Hopf problems which may be solved exactly~\citep{Daniele2014}.  Otherwise approximative methods are required, with for example Pad\'e approximants~\citep{abrahams1997solution} having been successfully applied to problems in elastodynamics~\citep{abrahams2000application}. Another popular approximative method in electromagnetism is the use of Fredholm factorisation~\citep{DanieleLombardi}.  Instead, here we adopt an iterative method first developed by \citet{Kisil2018}, intended to approximate the solution to $2\times 2$ matrix Wiener--Hopf problems with exponential factors, which has successfully been applied to problems in acoustics~\citep{KisilAyton2018}, and has subsequently been extended to $n\times n$ matrices by \citet{Priddin}, including discussions on how to implement such a procedure numerically.  The implementation requires numerical evaluations of Cauchy integrals (as in the scalar case), for which spectrally accurate numerical methods have been developed by \citet{slevinsky2017fast,olver2011computing,trogdon2015riemann}.  We note in passing that Wiener--Hopf problems bear a close relationship to Riemann-Hilbert problems~\citep{KisilRiemann} and so one may alternatively frame the problem as a Riemann-Hilbert problem and solve that problem numerically~\citep{trogdon2015riemann,Elena}, although this is not pursued further here.

One final complication of our contact problem is that the location of the contact region itself is unknown, and is require to be solved as part of the problem~\citep{Howell}.  Such free-boundary problems are inherently more complicated than comparable problems where the location of the boundary is fixed or is known a priori, and there is no generally applicable methodology for solving free-boundary problems~\citep{Howison}.  One typical approach in contact mechanics is to frame the problem as a variational inequality~\citep{Elliott,Fichera}.  An analytical method involving the use of Mellin transforms~\citep{Craster} has also been used in certain cases.  Here, we adopt an iterative procedure to re-estimate the contact region based on the solution using the previous estimate in order to ensure continuity of the solution as we transition from one region to the next.  This technique is specific to the Wiener--Hopf-based solution method used, which in general results in discontinuities at the transitions between boundary conditions.

A detailed description of the physical problem to be solved, together with its mathematical formulation, is given in section~\ref{sec:Formulate}, including the general solution for any boundary conditions in section~\ref{ssec:General}.  Considering the specific boundary conditions in different regions then leads to the construction of the matrix Wiener--Hopf problem in section~\ref{sec:WH}. This Wiener--Hopf problem is then solved using an iterative method in section~\ref{sec:Iterative}, including details of the numerical implementation of the solution method in section~\ref{ssec:Numerical}.  Details of the iterative solution to the resulting free-boundary problem are then presented in section~\ref{sec:Free-boundary}, illustrated with some numerical results.  The results of this analysis and numerics is presented in section~\ref{sec:Results} for a variety of parameters.  Finally, in section~\ref{sec:Conclusion}, conclusions are discussed along with avenues for potential future research.

\section{Mathematical formulation}\label{sec:Formulate}

\begin{figure}%
\centering%
\includegraphics{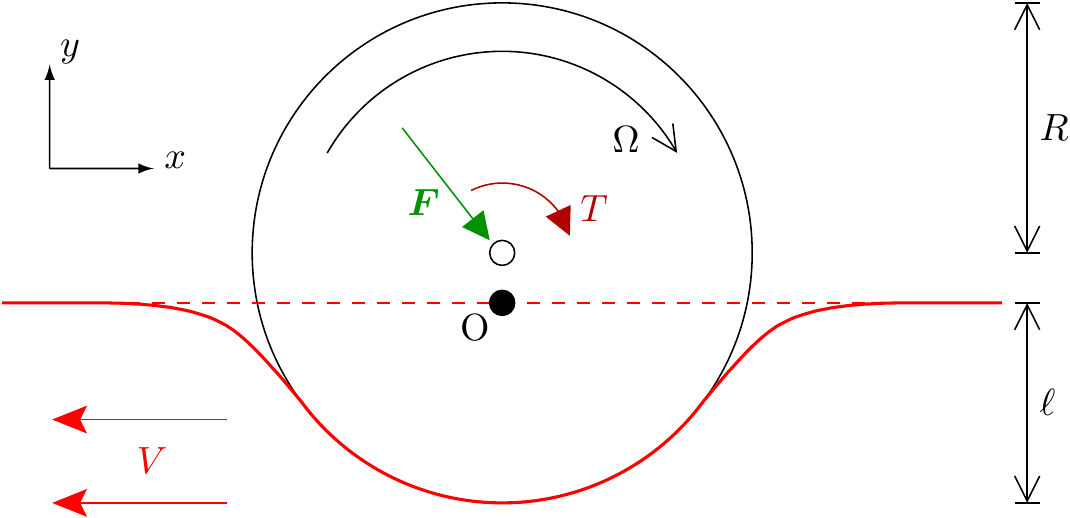}%
\caption{Schematic of a cylinder rolling along an elastic half space.  The cylinder, of radius $R$, moves at a linear velocity $V$ in the $x$-direction along the elastic half space.  The origin of the coordinate system (labelled~O) is taken in a frame of reference moving with the cylinder, directly below the centre of the cylinder at the height of the undeformed elastic surface. The cylinder is in contact with the elastic surface between the points $-a$ and $d$.  A force $\vect{F}$ and torque $T$ are applied to the centre of the cylinder, causing the cylinder to roll about its central axis with angular velocity $\Omega$ and be indented a depth $\ell$ into the elastic half-space.}
\label{fig:FSschematic1}
\end{figure}

We consider the situation shown schematically in figure~\ref{fig:FSschematic1}.
A cylinder of radius $R$ is pushed into an elastic half-space $y < 0$ with a force $\vect{F}$, resulting in contact with the elastic half-space for $-a\leq x\leq d$ and a maximum indentation of depth $\ell$.  A torque $T$ causes the cylinder to rotate across the half-space with an angular velocity $\Omega$, resulting in a translation in the $x$-direction at linear velocity $V$.  In what follows, we choose a frame of reference moving with the cylinder, such that the centre of the cylinder is located at $x=0$, and such that $y=0$ is the undeformed surface of the elastic half-space.

As described in appendix~\ref{app:governing-equations}, the full plain-strain nonlinear elastic governing equations are nondimensionalized, assuming elastic displacements are of order $O(\ell)$, velocities are of order $O(V)$, and coordinates $x$ and $y$ are of order $O(-a,d) = O(\sqrt{R\ell})$.  This results in a dimensionless parameter $\epsilon = \sqrt{\ell/R}$.  In nondimensional terms, we may take $R=1$, $V=1$, and displacement in the elastic half-space to be $\vect{u}(x,y) = (\epsilon u, \epsilon v)$.  Assuming that $\epsilon \ll 1$ results in the linear elastic governing equations
\begin{subequations}\begin{align}
\rho\!\left(\frac{\partial}{\partial t} -\frac{\partial}{\partial x}\right)^{\!\!2}\!\!u
&= (1+2\mu)\frac{\partial^2 u}{\partial {x}^2} + (1+\mu)\frac{\partial^2 v}{\partial x\partial y} + \mu\frac{\partial^2 u}{\partial {y}^2},\\
\rho\!\left(\frac{\partial}{\partial t} -\frac{\partial}{\partial x}\right)^{\!\!2}\!\!v
&= (1+2\mu)\frac{\partial^2 v}{\partial {y}^2} + (1+\mu)\frac{\partial^2 u}{\partial x\partial y} + \mu\frac{\partial^2 v}{\partial {x}^2},
\end{align}\label{equ:linearized-equations}\end{subequations}
where $\lambda$ and $\mu$ are the Lam\'e coefficients for the elastic material and both have been nondimensionalized such that $\lambda=1$.  The Cauchy stress tensor $\mat{\tau}$ is given in terms of $u$ and $v$ as
\begin{equation}
\mat{\tau} = \begin{pmatrix}
\tau_{xx} & \tau_{xy} & 0\\
\tau_{xy} & \tau_{yy} & 0\\
0 & 0 & \tau_{zz}
\end{pmatrix}
 = \begin{pmatrix}
(1+2\mu)\frac{\partial u}{\partial x} + \frac{\partial v}{\partial y} & \mu\!\left(\frac{\partial u}{\partial y} + \frac{\partial v}{\partial x}\right) & 0\\[1ex]
\mu\!\left(\frac{\partial u}{\partial y} + \frac{\partial v}{\partial x}\right) & \frac{\partial u}{\partial x} + (1+2\mu)\frac{\partial v}{\partial y} & 0\\[1ex]
0 & 0 & \left(\frac{\partial u}{\partial x} + \frac{\partial v}{\partial y}\right)
\end{pmatrix}.
\end{equation}
Note that the governing equations above are given in their more-general time-dependent form, despite the solution required being a time-invariant steady state.  The time-dependence will be used later in order to regularize the problem, and to ensure a causal solution is found.

\subsection{Boundary conditions}

As described above, the cylinder is in contact with the surface in the region $-a \leq x \leq d$. The points $-a$ and $d$ are unknown and are to be found as part of the solution.  For $x < -a$ and $x>d$, the surface is traction free, and the resulting linearized boundary conditions are given in~\eqref{app:equ:linearized-free} as
\begin{align}
\tau_{xy} = \tau_{yy} &= 0 && \text{for}\qquad y= 0, \qquad x < -a \text{ and } x > d.\label{equ:linearized-free}  
\end{align}

For $-a\leq x\leq d$, the elastic material is in contact with, but does not penetrate, the cylinder.  This results in the linearized boundary condition given in~\eqref{app:equ:linearized-contact},
\begin{align}
v &= \tfrac{1}{2}{x}^2 - 1, 
&& \text{for}\qquad y= 0, \qquad -a \leq x \leq d.
\label{equ:linearized-contact}\end{align}

We also have a traction boundary condition for $-a \leq x\leq d$.  Here, we consider the case of a cylinder which is slipping over the surface of the elastic half-space, either because the cylinder is rotating too rapidly (a ``wheel spin'' type condition) or too slowly (a ``locked brake'' type condition).  Assuming Coulomb friction with a friction coefficient $\mu_0$, the linearized boundary condition is given in~\eqref{app:equ:linearized-friction} as
\begin{align}
\tau_{xy} =\pm\mu_{0}\tau_{yy} && \text{for}\qquad y= 0, \qquad -a < x < d, \label{equ:linearized-friction}
\end{align}
where $\pm = \sgn(R\Omega/V-1)$ is given by the direction of slip.  The case that $R\Omega/V \approx 1$ would result in a stick--slip boundary condition, and is not considered further here.

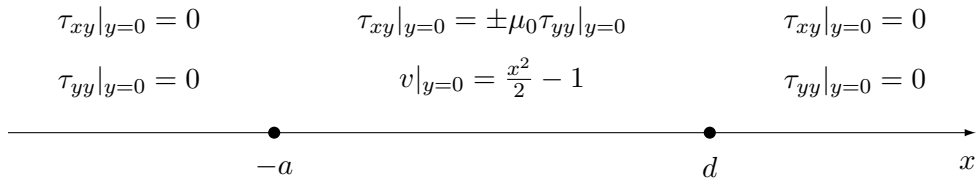
\begin{figure}
{\centering%
\setlength{\unitlength}{0.1\textwidth}%
\begin{picture}(8,2)
\put(1,1.3){\makebox(0,0)[b]{$\tau_{xy}|_{y=0}=0$}}
\put(4,1.3){\makebox(0,0)[b]{$\tau_{x y}|_{y=0} = \pm \mu_{0} \tau_{yy}|_{y=0}$}}
\put(7,1.3){\makebox(0,0)[b]{$\tau_{xy}|_{y=0}=0$}}
\put(1,0.8){\makebox(0,0)[b]{$\tau_{yy}|_{y=0}=0$}}
\put(4,0.8){\makebox(0,0)[b]{$v|_{y=0} = \frac{x^2}{2} - 1$}}
\put(7,0.8){\makebox(0,0)[b]{$\tau_{yy}|_{y=0}=0$}}
\put(2.2,0.3){\makebox(0,0)[t]{$-a$}}
\put(5.8,0.3){\makebox(0,0)[t]{$d$}}
\put(0,0.5){\vector(1,0){8}}
\put(8,0.3){\makebox(0,0)[rt]{$x$}}
\put(2.2,0.5){\circle*{0.1}}
\put(5.8,0.5){\circle*{0.1}}
\end{picture}\par}%
  \caption{The linearized boundary conditions and the regions where they hold.}\label{fig:FSbc}
\end{figure}

These boundary conditions are summarized in figure~\ref{fig:FSbc}.
The resultant mathematical problem is a mixed-boundary problem, with two boundary conditions in each region.  Moreover, the junctions between these regions are the free-boundary points $-a$ and $d$, whose position must be determined as part of the solution.  In what follows, the mixed-boundary problem is reformulated as a matrix Wiener--Hopf problem.  The free-boundary points are then located a posteriori based on an iterative method, by requiring continuity of the solution at these free-boundary points.

\subsection{General solution}\label{ssec:General}

The governing equations~\eqref{equ:linearized-equations} above are given in time-dependent form.  In what follows, we assume a time dependence of the form $\exp\{\I\omega t\}$.  Since we are interested in the steady-state solution, we will eventually be interested in the limit $\omega\to0$, although assuming nonzero $\omega$ will be seen to give greater regularity to the intermediate solutions.  Moreover, in what follows we take this limit assuming that $\Imag(\omega)<0$, so that the causal solution is given by assuming the far-field boundary conditions that $u\to 0$ and $v\to 0$ at infinity.  For any variable $\phi(x,y,t)$, we therefore set $\phi(x,y,t) = \Real(\hat{\phi}(x,y)\exp\{\I\omega t\})$.  Moreover, provided that $|\hat{\phi}|\to 0$ as $|x|\to\infty$, we may Fourier transform in the $x$-direction,
\begin{align}
\tilde{\phi}(k,y)&= \int^{\infty}_{-\infty} \hat{\phi}(x,y) \e^{\I k x}\,\intd x &
&\Leftrightarrow&
\phi(x,y,t) &= \Real\left(\frac{1}{2\pi}\int^{\infty}_{-\infty} \tilde{\phi}(k,y) \exp\{\I\omega t-\I kx\}\,\intd k\right).
\end{align}
Applying Fourier transforms to the governing equations~\eqref{equ:linearized-equations}, we find that

\begin{align} 
\mu \frac{\partial^2 \tilde{u} }{\partial y^2} - ik (\lambda + \mu) \frac{\partial \tilde{v}}{\partial y} + (\rho V^2 (\omega + k)^{2} - (\lambda +2 \mu)k^2)\tilde{u} = 0,\\
(\lambda +2 \mu) \frac{\partial^2 \tilde{v} }{\partial y^2} - ik (\lambda + \mu) \frac{\partial \tilde{u}}{\partial y} + (\rho V^2 (\omega + k)^{2} -  \mu k^2)\tilde{v} = 0.
\end{align}
To enable a physically attainable solution deformations must decay far from the roller so an ansatz which decays in the $y$-direction is apt. A solution may be found by taking the following ansatz,
\begin{equation}
\tilde{u}(k,y) = A(k)e^{\gamma(k)y}, \quad \tilde{v}(k,y) = A(k)B(k)e^{\gamma(k)y},
\end{equation}
where $A(k),B(k)$ and $m(k)$ are all unknowns. Applying the ansatz to the transformed governing equation gives a fourth order polynomial to solve, where the solution gives
the following general solution to the governing equation
\begin{align}
u(x,y)= \frac{1}{{2\pi}}\int^{\infty}_{-\infty} \bigg[ A_{1}(k)e^{y \gamma_{1}(k)} &+ A_{2}(k)e^{y\gamma_{2}(k)} \bigg] e^{-ik x} dk, \\
v(x,y)= \frac{1}{{2\pi}}\int^{\infty}_{-\infty} \bigg[ A_{1}(k)B_{1}(k)e^{y \gamma_{1}(k)}& + A_{2}(k) B_{2}(k)e^{y \gamma_{2}(k)} \bigg] e^{-ik x} dk.
\end{align}
Due to the decay in the elastic media only two terms of the solutions to the fourth order polynomial remain, $\gamma_1(k)$ and $\gamma_2(k)$. The functions $ \gamma_1(k), \gamma_2(k), B_1(k)$ and $B_2(k)$, are now known and shown below
\begin{align}
\gamma_{1}(k) =&
\sqrt{k^2 - \frac{\rho V^2}{\lambda + 2\mu}\big(\omega+k\big)^2}
,
&&\gamma_{2}(k) =
\sqrt{k^2 - \frac{\rho V^2}{\mu}\big(\omega+k\big)^2}\notag\\
\\
B_{1}(k) =& { \frac{\I\gamma_1(k)}{k}} = \sqrt{\frac{\rho V^2 (1 + \frac{\omega}{k})^2}{\lambda + 2 \mu}-1},
 &&B_{2}(k) = { \frac{\I k}{\gamma_2}} = -\frac{1}{\sqrt{\frac{\rho V^2 (1 + \frac{\omega}{k})^2}{\mu}-1}}.
\end{align}
The branch cuts of $\gamma_1$ and $\gamma_2$ should be chosen such that $\mathrm{real}(\gamma) > 0$, with the branch cuts of $B_1$ and $B_2$ being chosen accordingly.  The only two unknowns remaining, $A_1 (k)$ and $A_{2} (k)$, remain be found by considering the boundary conditions.

By considering the asymptotic behaviour as $ k \rightarrow 0 $, it becomes apparent that the structure of the general solution presented above is incorrect. The general solution fails due to the asymptotic behaviour of $B_1(k)$, which tends to infinity as $k$ approaches zero. To regularise the problem one may redefine in the following way,
\begin{align} \label{eqn: FSregularised solution}
  u(x,y)=& \frac{1}{{2\pi}}\int^{\infty}_{-\infty} \bigg[ \hat{A_{1}}(k) \hat{B_1}(k) e^{y \gamma_{1}(k)} + A_{2}(k)e^{y\gamma_{2}(k)} \bigg] e^{-ik x} dk, \\
  v( x,y)=& \frac{1}{{2\pi}}\int^{\infty}_{-\infty} \bigg[ \hat{A_{1}}(k)e^{y \gamma_{1}(k)} + A_{2}(k) B_{2}(k)e^{y \gamma_{2}(k)} \bigg] e^{-ik x} dk.
\end{align}
Explicitly, this regularization is defining the functions $\hat{A_{1}}(k)$ and $\hat{B_1}(k)$ as
\begin{align}
  &\hat{A_{1}}(k) = A_{1}(k)B_{1}(k), 
  &&\hat{B_{1}}(k) = \frac{1}{B_1 (k) } .
\end{align}
Observe that $\hat{B_1}(k) \rightarrow 0 $ as $k \rightarrow 0$, which simplifies the general solution into a form which agrees with the $k \rightarrow 0$ limit of the fourth order polynomial.

\section{Constructing the matrix Wiener--Hopf equation}\label{sec:WH}

The general solution was found by Fourier transforming the governing equations and solving a fourth order polynomial in the spectral domain, to make further progress, information from the mixed boundary values is required. The transformation of the mixed boundary value problem is troublesome due to the intervals of the spatial domain that the boundary conditions exist over, hence half-range transforms are used and defined in section \ref{ssec:HR}.

The transformed boundary conditions may be assembled into a format where a Wiener--Hopf technique may be applied. In this case a matrix Wiener--Hopf equation is formed, where only very few exact solutions exist. Careful construction of the matrix Wiener--Hopf equation ensures that a format is assembled where the iterative method by \citet{Priddin} may be applied.

\subsection{Half-range transforms}\label{ssec:HR}

To derive the unknown functions $\hat{A_{1}}(k)$ and $A_{2}(k)$ the boundary conditions must be transformed. For this purpose, the boundary conditions are transformed  over their respective regions, where the finite-range transforms are written as the difference of various half-range transforms. Then  the transformed boundary conditions are assembled into a matrix Wiener--Hopf problem.

We define a half-range Fourier transforms with respect to the point $x = L$  by considering the full-range Fourier transform and splitting as follows,
\begin{align} \label{eqn: FSCartPap1}
\widetilde{\phi}^{L}(k , y) =& \int^{\infty}_{-\infty}  \phi(x, y) e^{ik(x-L)} dx  \\ \nonumber
		  =&  \int^{\infty}_{L}  \phi(x, y) e^{ik(x-L)} dx &&+  \int^{L}_{-\infty}  \phi(x, y) e^{ik(x-L)} dx  \\ \nonumber
		  =& \qquad  \widetilde{\phi}^{L}_{+}(k , y)  &&+ \qquad \widetilde{\phi}^{L}_{-}(k , y).
\end{align}
With $\widetilde{\phi}^{L}_{+}(k , y)$ analytic in the upper half of the complex $k$--plane and $\widetilde{\phi}^{L}_{-}(k , y)$ analytic in the lower half. A half-range Fourier transform is defined for each of the junction points of the boundary conditions, $ x = -a,  d$. The shifted full-range transforms may be related to one centred at $ x = 0$ by
\begin{equation}
\widetilde{\phi}^{L}(k , y)e^{ikL}  = \widetilde{\phi}(k , y).
\end{equation}
Finally, the transformation of a finite interval may be related to half-range transforms in the following way
\begin{align}
 \widetilde{\phi}^{[L_{1},L_{2}]}(k,y) &=  \int_{L_{2}}^{L_{1}} \phi(x,y) e^{ik x} dx ,\\
				  &= \widetilde{\phi}^{L_{1}}_{-}(k , y)e^{ikL_{1}} - \widetilde{\phi}^{L_{2}}_{-}(k , y)e^{ikL_{2}}
				  \\
				  &= \widetilde{\phi}^{L_{2}}_{+}(k , y)e^{ikL_{2}} - \widetilde{\phi}^{L_{1}}_{+}(k , y)e^{ikL_{1}}.
\end{align}
This gives the tools to transform the boundary conditions over their respective regions. 

\subsection{Transformation of boundary conditions}\label{ssec:Transform}

To construct of the simplest matrix Wiener--Hopf equation, it is necessary to consider the boundary conditions carefully. In fact, by rearranging the boundary conditions, the size of the matrix Wiener--Hopf may be halved. We shall transform each boundary condition over their respective regions and assemble them to form a matrix Wiener--Hopf equation, with a structure suitable for the iterative method may be applied to. To enable the application of the iterative method \cite{Priddin}, the matrices are required to be triangular and have the correct analyticity of the exponential terms.

\begin{figure}
%   \centering
  {\centering%
\setlength{\unitlength}{0.1\textwidth}%
\begin{picture}(8,2)
% \put(1,1.3){\makebox(0,0)[b]{$\tau_{xy}|_{y=0}=0$}}
\put(4,1.3){\makebox(0,0)[b]{$\tau_{x y}|_{y=0} \mp \mu_{0} \tau_{y y}|_{y=0} = 0$}}
% \put(7,1.3){\makebox(0,0)[b]{$\tau_{xy}|_{y=0}=0$}}
\put(1,0.8){\makebox(0,0)[b]{$\tau_{yy}|_{y=0}=0$}}
\put(4,0.8){\makebox(0,0)[b]{$v|_{y=0} = \frac{x^2}{2} - 1$}}
\put(7,0.8){\makebox(0,0)[b]{$\tau_{yy}|_{y=0}=0$}}
\put(2.2,0.3){\makebox(0,0)[t]{$-a$}}
\put(5.8,0.3){\makebox(0,0)[t]{$d$}}
\put(0,0.5){\vector(1,0){8}}
\put(5.6,1.4){\vector(1,0){2.4}}
\put(2.4,1.4){\vector(-1,0){2.4}}
\put(8,0.3){\makebox(0,0)[rt]{$x$}}
\put(2.2,0.5){\circle*{0.1}}
\put(5.8,0.5){\circle*{0.1}}
\end{picture}\par}%
  \caption{The figure shows another way to represent the full-slip boundary conditions, minimising the number of junctions between individual boundary conditions. This rearrangement reduces the number of scattering points (junctions) and so the size of the matrix Wiener--Hopf.} \label{fig: FSbc slip}
\end{figure}
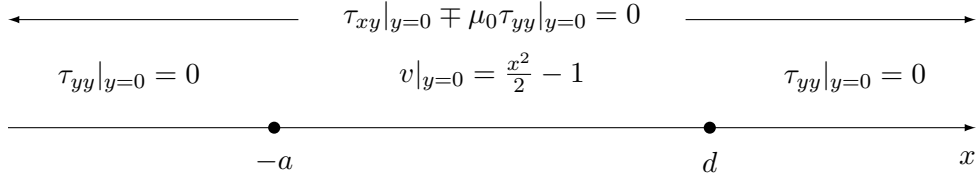

The first strategy is to minimise the number of junctions between boundary conditions. The key reduction is to rewrite the stress-free boundary conditions to include the slipping boundary condition, referring to the top of figure \ref{fig: FSbc slip}. Recall the stress-free boundary conditions and manipulate to find
\begin{align}
    \tau_{x y}|_{y=0} = \tau_{y y}|_{y=0 } = 0  \iff    \tau_{x y}|_{y=0} \mp \mu_0 \tau_{y y}|_{y=0 } = \tau_{y y}|_{y=0 } =  0 .
\end{align}
So now the slip boundary condition
holds across the entire surface of the half-space and 
\begin{equation} \tau_{yy}|_{y=0} = 0  \label{eqn: FSnormal stress}\end{equation} 
holds in in the free-boundary region $x < -a$ and $x > d$. This rearrangement reduces the number of junctions from four to two which gives a $(2 \times 2)$ matrix Wiener--Hopf problem.

Then by taking a full-range transform of the slipping boundary condition, the following is derived
\begin{align}\label{eqn: FS A1 A2}
\widetilde{{\tau_{x y}|_{y=0}\mp\mu_{0} \tau_{yy}|_{y=0}}} &= 0
\Rightarrow  \hat{A_{1}}(k) = - \frac{m_{2}^{\mp}(k)}{m_{1}^{\mp}(k)}  A_{2}(k) = n(k) A_{2}(k).
\end{align}
With $m_{i}^{\mp}$ a known function, defined from the full-range transforms \eqref{eqn: FStmp2}. The rearrangement of the boundary conditions leads to a relationship between $\hat{A_1}(k)$ and $A_2(k)$, which halves the number of unknowns in the system. In this instance reducing the number of junctions reduces the number of unknowns which leads to a reduction in the size of the matrix Wiener--Hopf equation.

To derive the Wiener--Hopf problem the unknown variables will be defined from the half-range transforms of $ v|_{y=0} $ and $\tau_{yy}|_{y=0}$. By considering the boundary conditions \eqref{equ:linearized-contact} and \eqref{eqn: FSnormal stress}, then taking a full-range transform to find the following
\begin{align} \label{eqn: FSshear stress free + slip}
\widetilde{{{\tau_{yy}}_{y=0}}} = e^{ikd}\wft{{{\tau_{yy}}_{y=0}}}{d}{}{-}= e^{-ika}\wft{{{\tau_{yy}}_{y=0}}}{-a}{}{+} , \\ \label{eqn: FScontact wh}
\widetilde{v} = e^{ikd}\wdft{v}{d}{}{+} + e^{-ika}\wdft{v}{-a}{}{-} + f(k).
\end{align}
Where the function $f(k)$\footnote{Where we find $f(k)$ by considering the finite range transform
\begin{align}
 f(k) &:= \int^{d}_{-a} \left( \frac{x^2}{2} - 1 \right) e^{ikx} d x \\
               &=  \frac{e^{-i k a}}{ik}\left( 1 - \frac{a^2}{2} - \frac{a}{i k } + \frac{1}{k^2 }  \right) - \frac{e^{i k d}}{ik}\left( 1 - \frac{d^2}{2} + \frac{d}{i k} + \frac{1}{k^2}  \right).
\end{align}
} has been defined to simplify the algebra. These two scalar Wiener--Hopf equations shall then be manipulated to form the matrix Wiener--Hopf equation by eliminating the full-range transform variables on the LHS, $\widetilde{{{\tau_{yy}}_{y=0}}}$ and $\widetilde{v}$.

In order to assemble the Wiener--Hopf matrix, the relationship between the two full-range transform variables $\widetilde{{{\tau_{yy}}_{y=0}}}$ and $\widetilde{v}$ must be found. We may find this relationship by considering all of the full-range transformations in terms of $\hat{A_1}(k)$ and $A_2(k)$,
\begin{align} \label{eqn: FSty}
\widetilde{{\tau_{yy}|_{y=0}}} =& \Big[(\lambda + 2 \mu ) \gamma_{1} - ik\lambda \hat{B_{1}} \Big] \hat{A_{1}} \\ 
 &+ \Big[(\lambda + 2 \mu ) B_{2} \gamma_{2} - ik\lambda \Big] A_{2} \\
=& n_{1}(k)\hat{A_{1}} + n_{2}(k)A_{2}, \\  \label{eqn: FSv}
\widetilde{v|_{y=0}} =& \hat{A_{1}} + B_{2}(k) A_{2},\\  \label{eqn: FStmp}
\widetilde{{\tau_{x y}|_{y=0}\mp\mu_{0} \tau_{yy}|_{y=0}}} =& \Bigg[\mu ( \gamma_{1}\hat{B_{1}} -ik ) \mp \mu_{0} \Big( ( \lambda + 2 \mu )\gamma_{1} - \hat{B_{1}} ik\lambda \Big) \Bigg] \hat{A_{1}} \\
&+ \Bigg[ \mu ( \gamma_{2} -ik  B_{2} ) \mp\mu_{0} (\Big( \lambda+ 2 \mu )  B_{2} \gamma_{2} - ik\lambda \Big) \Bigg] A_{2}\\
=& m_{1}^{\mp}(k) \hat{A_{1}} + m_{2}^{\mp}(k) A_{2}. \label{eqn: FStmp2}
\end{align}
Where the functions $n_1 (k) , n_2 (k) , m_{1}^{\mp}(k) $ and $m_{2}^{\mp}(k)$ are defined by the above.

To assemble the matrix, the unknown functions $\hat{A_{1}}(k)$ and $A_2(k)$ are eliminated from the equations \eqref{eqn: FS A1 A2},\eqref{eqn: FSty} and \eqref{eqn: FSv} to form equations where the unknowns are solely full-range transformed variables. The full-range transformed variables may then be replaced with their half-range counterparts from \eqref{eqn: FSshear stress free + slip} and \eqref{eqn: FScontact wh}, and then assembled into a matrix Wiener--Hopf equation with the correct structure for the iterative method. The rearranging of the stress-free boundary conditions has reduced the number of junctions by half, which in turn reduces the complexity of the Wiener--Hopf from a $(4 \times 4)$ matrix to a $( 2 \times 2)$ matrix. Finally we arrive at the following ($2 \times 2 $) matrix Wiener--Hopf equation,
\begin{multline} \label{eqn: FSmatrix WH 1}
\begin{pmatrix}
    1 &  0 \\
    \\ 
    - e^{ik(a+d)} & K(k) 
    \end{pmatrix}  
 \begin{pmatrix}
    \wdft{v}{d}{}{+}(k)  \\
    \\
    \wft{{{\tau_{yy}}_{y=0}}}{-a}{}{+}(k)
  \end{pmatrix}
  \\
  =
  \begin{pmatrix}
    K(k) & -e^{-ik(a+d)} \\
     &   \\
     0	 & 1
      \end{pmatrix} 
  \begin{pmatrix}
    \wft{{{\tau_{yy}}_{y=0}}}{d}{}{-}(k)  \\
    \\
    \wdft{v}{-a}{}{-}(k)
  \end{pmatrix}
   +
   \begin{pmatrix}
    -e^{-ikd}f(k) \\
    \\
     e^{ika}f(k)
\end{pmatrix}.
\end{multline}
Where $K(k)$ is found by rearranging the transformed boundary conditions and is
\begin{align}
{K}(k) &= \frac{[ n(k) +  B_{2}(k) ]}{[n_{1}(k) n(k) + n_{2}(k)] },& \text{ with } {K}(k) &= O\left(\frac{1}{|k|}\right) \text{ as } |k| \rightarrow \infty.
\end{align}

It is now possible to apply the iterative method to the matrix Wiener--Hopf problem as the exponential terms are of the required analyticity and the matrices have the required structure. The unknowns in the problem are $\wdft{v}{d}{}{+}(k)$ and $ \wft{{{\tau_{yy}}_{y=0}}}{-a}{}{+}(k)$ analytic in the upper half-plane, with $ \wft{{{\tau_{yy}}_{y=0}}}{d}{}{-} (k)$  and $\wdft{v}{-a}{}{-}(k)$ analytic in the lower half-plane.

{To ensure that the least singular solutions are found, it is required that $\wdft{v}{d}{}{+}(k), \wft{{{\tau_{yy}}_{y=0}}}{-a}{}{+}  $ and $\wft{{{\tau_{yy}}_{y=0}}}{d}{}{-} , \wdft{v}{-a}{}{-}(k) $ decay to $0$ as $|k| \rightarrow \infty $ in the upper and lower half-planes respectively. This behaviour may be ensured by imposing the edge conditions for the stress
\begin{align}
    {\tau_{yy}}_{y=0}(x) \rightarrow 0 , &\text{ as } x \rightarrow -a^{+}, &&{\tau_{yy}}_{y=0}(x) \rightarrow 0 , \text{ as } x \rightarrow d^{-}.
\end{align}
These two edge conditions ensure that the variables $\wft{{{\tau_{yy}}_{y=0}}}{-a}{}{+}$ and $\wft{{{\tau_{yy}}_{y=0}}}{d}{}{-}$ decay quadratically. Whereas the edge conditions for the vertical displacement
\begin{align}
    {v}_{y=0}(x) \rightarrow \frac{a^2}{2} -1  , &\text{ as } x \rightarrow -a^{-},  
    &&{v}_{y=0}(x) \rightarrow \frac{d^2}{2} -1 , \text{ as } x \rightarrow d^{+},
\end{align}}
ensures linear decay for $\wdft{v}{d}{}{+}(k)$ and $\wdft{v}{-a}{}{-}(k).$

\section{Application of the iterative method}\label{sec:Iterative}

The mathematical formulation of the physical problem has been derived and recast as a matrix Wiener--Hopf problem. The presence of the exponential factors in the matrix Wiener--Hopf \eqref{eqn: FSmatrix WH 1} suggests the use of the approximative factorisation method developed by \citet{Kisil2018}, extended by \citet{Priddin}, and applied in acoustics by \citet{KisilAyton2018}. The arrangement of the Wiener--Hopf problem into triangular matrices gives further structure to enable the factorisation more easily. The method considers a matrix Wiener--Hopf such as \eqref{eqn: FSmatrix WH 1} and approximates the exponential terms to zero. Once an initial approximation has been made the matrix Wiener--Hopf may be considered as a series of scalar Wiener--Hopf equations, in these scalar equations one may use additive and multiplicative decompositions to arrive at a form in which Liouville's Theorem may be applied. The approximation is corrected for by iterating through the scalar equations with the exponential terms reintroduced, and additively decomposing the correction term into a form where Liouville's may be applied again.

\begin{figure}
  \centering\begin{tikzpicture}
    \node[anchor=south west,inner sep=0] at (0,0) { \includegraphics[width=5in]{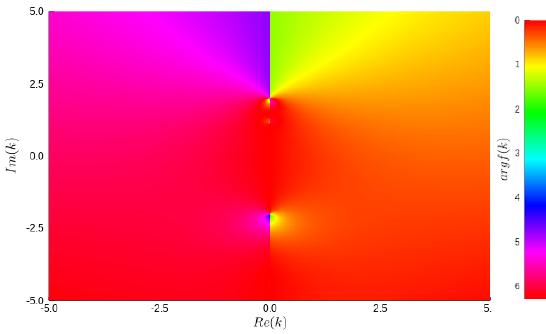}    };
    \draw (10,2.4)node[above] {$\mathcal{D}_-$} 
    (6.3,4.2)node[below] {$\mathcal{D}$}
    (2,5)node[above] {$\mathcal{D}_+$}
    (4,4.75)node[below] {$\Gamma_-$}
    (8,3.05)node[above] {${\Gamma}_+$};
    \draw [decorate,decoration={brace,amplitude=14pt},xshift=-4pt,yshift=0pt](3,3) -- (3,7.5);
    \draw [decorate,decoration={brace,amplitude=14pt,mirror,raise=4pt},yshift=0pt](9,0.8) -- (9,4.8) ;
    \draw[thick] (1.2,4.8) -- (11.4,4.8);
    \draw[thick] (1.2,3) -- (11.4,3);
\end{tikzpicture}
  \caption{This figure shows a phase-portrait for $K(k)$ with $\lambda = 210000, \mu =  81000, \mu_0 = 0.3, \rho = 7850, V = 2$ and $\omega = -5i$. The introduction of $\omega$ gives the effect of pulling the branch cut open to allow a strip of analyticity through, $\mathcal{D}$. One may take an inversion contour near the bottom of the strip to justify the approximation of $e^{-ik(a+d)}$ to zero, the contours  for $\Gamma_+$ and $\Gamma_-$ illustrates this.}
  \label{fig: FSadditive}
\end{figure}
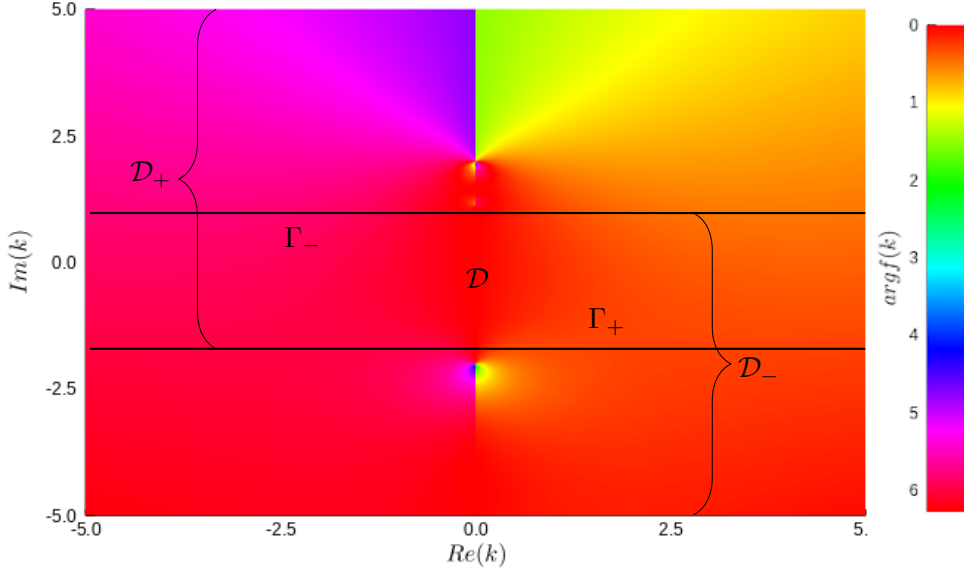

To signify that the process is iterative, we introduce the notation $\wft{\phi}{L}{n}{\pm}$ as the n-th iteration of $\wft{\phi}{L}{}{\pm}$. 
Recalling \eqref{eqn: FSmatrix WH 1} and taking the initial estimate,
\begin{equation} \label{eqn: FSmatrix WH 2}
  \begin{pmatrix}
    1 &  0& \\
    \\ 
    0  & K(k) &  
    \end{pmatrix}  
   \begin{pmatrix}
     \wdft{v}{d}{0}{+}  \\
     \\
     \wft{{{\tau_{yy}}_{y=0}}}{-a}{0}{+}
   \end{pmatrix}
  =
  \begin{pmatrix}
   & K(k) & 0  &\\
         \\
   & 0	 & 1 &
      \end{pmatrix} 
   \begin{pmatrix}
     \wft{{{\tau_{yy}}_{y=0}}}{d}{0}{-}  \\
     \\
     \wdft{v}{-a}{0}{-}
   \end{pmatrix}
   +
   \begin{pmatrix}
    -e^{-ikd}f(k) \\
    \\
     e^{ika}f(k)
\end{pmatrix}.
\end{equation}
The justification for making such an approximation stems from the choice of inverse contours one may take in the overlapping strip of analyticity $\mathcal{D}$, as demonstrated in figure \ref{fig: FSadditive}. As one may take an inversion contour anywhere within the strip in figure \ref{fig: FSadditive}, choosing a contour near the bottom of the strip would set $e^{-ik(a+d)}$ to be close to zero, like wise an inversion contour near the top of the strip would set $e^{ik(a+d)}$ to be close to zero, hence justifying the approximation. 

One may see that the top row of the approximated matrix Wiener--Hopf problem may now be solved, subject to decompositions. Considering each of the scalar Wiener--Hopf equations from the matrix separately gives,
\begin{align} \label{eqn: FSscalar WH 1}
  \wdft{v}{d}{0}{+}  &= K(k) \wft{{{\tau_{yy}}_{y=0}}}{d}{0}{-}  - e^{-ikd}f(k),\\ \label{eqn: FSscalar WH 2}
 K(k)\wft{{{\tau_{yy}}_{y=0}}}{-a}{0}{+} &=  \wdft{v}{-a}{0}{-} +  e^{ika}f(k).
\end{align} 
Referring to \eqref{eqn: FSscalar WH 1}, one may take a multiplicative decomposition of $K(k)$\footnote{Define the notation for additive and multiplicative decompositions. Let superscripts
$F(k) = F^+(k) + F^-(k)$ signify an additive decomposition whereas subscripts $F(k) = F_+(k) F_-(k)$  signify a multiplicative decomposition. Where the decomposition are analytic in either the upper ($+$) or lower ($-$) half of the complex $k$-plane. } and divide through by ${K}_+(k)$ and then apply an additive decomposition of the resulting forcing term giving
\begin{equation} \label{eqn: FSscalar WH 3}
  \frac{1}{{K}_+(k)}  \wdft{v}{d}{0}{+}  = {{K}_-(k)} {\wft{{{\tau_{yy}}_{y=0}}}{d}{0}{-}} - \left[\frac{1}{{K}_+(k)} e^{-ikd}f(k)\right]^+ - \left[\frac{1}{{K}_+(k)}e^{-ikd}f(k)\right]^-. 
\end{equation} 
The regions of analyticity may be extended to the entire complex plane via analytic continuation~\citep{Noble}, so we may introduce an entire function $J(k)$ such that,
\begin{equation} \
  J(k) =  \frac{1}{{K}_+(k)}  \wdft{v}{d}{0}{+}  +   \left[\frac{1}{{K}_+(k)} e^{-ikd}f(k)\right]^+  = {{K}_-(k)} {\wft{{{\tau_{yy}}_{y=0}}}{d}{0}{-}}   -  \left[\frac{1}{{K}_+(k)} e^{-ikd}f(k)\right]^-. 
\end{equation} 
Liouville's theorem may be applied to the above since we ensured $\wft{{{\tau_{yy}}_{y=0}}}{d}{0}{-} \sim k^{-2} $ as $k \rightarrow \infty$ and the exponential decay of the forcing which gives $J(k) = 0$. Applying Liouville's theorem leads to the initial approximative solutions,
\begin{align} 
  \wdft{v}{d}{0}{+}(k) =  - {K}_+(k){ \left[\frac{1}{{K}_+(k)} e^{-ikd}f(k)\right]^+},& \quad 
  \wft{{{\tau_{yy}}_{y=0}}}{d}{0}{-}(k) = \frac{\left[\frac{1}{{K}_+(k)} e^{-ikd}f(k)\right]^-} {{K}_-(k)}. 
\end{align}
Applying the same argument for equation \eqref{eqn: FSscalar WH 2} leads to the following initial solutions,
\begin{align} \label{eqn: FSsol v - 0}
  \wdft{v}{-a}{0}{-}(k) =  -{{K}_-(k)}  {\left[\frac{1}{{K}_-(k)}e^{ika}f(k)\right]^-},& \quad 
  \wft{{{\tau_{yy}}_{y=0}}}{-a}{0}{+}(k) =\frac{1}{{K}_+(k)}  \left[\frac{1}{{K}_-(k)}e^{ika}f(k)\right]^+. 
\end{align}

Now we refer back to \eqref{eqn: FSmatrix WH 1}, and reintroduce the exponential terms, 
\begin{multline} \label{eqn: FSmatrix WH 3}
  \begin{pmatrix}
    1 &  0& \\
    \\ 
    -e^{ik(a+d)}  & K(k) &  
    \end{pmatrix}  
   \begin{pmatrix}
     \wdft{v}{d}{0}{+}  \\
     \\
     \wft{{{\tau_{yy}}_{y=0}}}{-a}{0}{+}
   \end{pmatrix}
\\
  =
  \begin{pmatrix}
   & K(k) &  - e^{-ik(a+d)} \\
         \\
   & 0	 & 1
      \end{pmatrix} 
   \begin{pmatrix}
     \wft{{{\tau_{yy}}_{y=0}}}{d}{0}{-}  \\
     \\
     \wdft{v}{-a}{0}{-}
   \end{pmatrix}
   +
   \begin{pmatrix}
    -e^{-ikd}f(k) \\
    \\
    e^{ika}f(k)
\end{pmatrix}.
\end{multline}
Where the initial solutions are known and we may continue to use the Wiener--Hopf technique iteratively, so we define the n-th iteration of the equations below. As in the the initial approximation, consider each row of the matrix Wiener--Hopf problem as scalar equations once again,
\begin{gather} \label{eqn: FSscalar WH 4}
    \wdft{v}{d}{n}{+}  =K(k) \wft{{{\tau_{yy}}_{y=0}}}{d}{n}{-} -  e^{-ik(a+d)} \wdft{v}{-a}{n-1}{-} -e^{-ikd}f(k), \\ \label{eqn: FSscalar WH 5}
    K(k)\wft{{{\tau_{yy}}_{y=0}}}{-a}{n}{+} - e^{ik(a+d)} \wdft{v}{d}{n-1}{+} =  \wdft{v}{-a}{n}{-}  +   e^{ika}f(k).
\end{gather} 

Then one may simply iterate through these equations, updating the n-th iteration with the solutions from the n-1-th iteration. Explicitly, the nth iteration equations will be
\begin{align}
  \wdft{v}{d}{n}{+} =&  - {K}_+(k){\left[\frac{1}{{K}_+(k)}e^{-ikd}f(k)\right]^+}  - {K}_+(k){\left[\frac{1}{{K}_+(k)}e^{-ik(a+d)} \wdft{v}{-a}{n-1}{-}\right]^+}, \\ 
  \wft{{{\tau_{yy}}_{y=0}}}{d}{n}{-} =&  \frac{\left[\frac{1}{{K}_+(k)}e^{-ikd}f(k)\right]^-}{{K}_-(k)} + \frac{\left[\frac{1}{{K}_+(k)}e^{-ik(a+d)} \wdft{v}{-a}{n-1}{-}\right]^-}{{K}_-(k)}, \\
  \wdft{v}{-a}{n}{-} =&  - {K}_-(k){\left[\frac{1}{{K}_-(k)}e^{ika}f(k)\right]^-} - {K}_-(k){\left[\frac{1}{{K}_-(k)}e^{ik(a+d)} \wdft{v}{d}{n}{+}\right]^-}, \\ 
  \wft{{{\tau_{yy}}_{y=0}}}{-a}{n}{+} =& \frac{ \left[\frac{1}{{K}_-(k)}e^{ika}f(k)\right]^+}{{K}_+(k)} + \frac{\left[\frac{1}{{K}_-(k)}e^{ik(a+d)} \wdft{v}{d}{n}{+}\right]^+}{{K}_+(k)}. 
\end{align}
The overview of the iterative method has omitted details on the computation of the upper or lower decompositions, which shall be examined in the next section.
To find the solution all that one needs to compute is the decompositions, however, to implement this in a practical sense it is best to proceed numerically.

\subsection{Numerical implementation}\label{ssec:Numerical}

To implement the iterative method it is necessary to develop a numerical approach to accurately compute the decompositions. Additive decomposition may be computed by considering Cauchy transforms~\citep{Daniele2014, Noble}, for $k\in \mathbb{C}$ and $F(k)$ entire
\begin{align}
  F(k)   &= \frac{1}{2 \pi i}\int_{\Gamma_+} \frac{F(x)}{x - k} dx  - \frac{1}{2 \pi i}\int_{\Gamma_-} \frac{F(x)}{x - k} dx \\
       &= \mathcal{C}_{\Gamma_+}[F](k) - \mathcal{C}_{\Gamma_-}[F](k) \\
       &= F^+(k) + F^-(k) .
\end{align}
Where $ k\in \mathcal{D}$ and the contours $\Gamma_+$ and $\Gamma_-$ are below and above $k$ respectively, as illustrated in figure \ref{fig: FSadditive}. The branch cuts have a strip of analyticity between them, $\mathcal{D}$, where the matrix Wiener--Hopf problem is analytic. It is further required that $F(k) \rightarrow 0$ as $|k| \rightarrow \infty$ for the Cauchy transform to hold.

The multiplicative factorisation of a function may be derived through the Cauchy transform by use of the logarithm. Again, for an entire complex function, $F (k) $,
\begin{equation}
  F(k)= e^{ [\log(F)]^+ + [\log(F)]^- }
       = F_+(k) F_-(k) .
\end{equation}
Where the additive decomposition of $\log(F)$ may be found via the Cauchy transform outlined previously. The requirement of the asymptotic behaviour of $F$ in the multiplicative factorisation is $F(k) \rightarrow 1$ as $|k| \rightarrow \infty$. Hence $\log(F) \rightarrow 0$ as $|k| \rightarrow \infty$, this asymptotic behaviour may be generated by normalising $F(k)$ by a known entire function $\gamma(k)$, which can be multiplicatively decomposed analytically~\citep{Noble}.

A natural question one asks is how to accurately and efficiently compute Cauchy transforms numerically. Traditionally one may use a quadrature rule, which approximates the branch cut due to the contour $\Gamma_{\pm}$ by a series of poles. This means the quadrature rule would give higher errors near to the contour, so one may deform the contour in line with Cauchy's integral theorem~\citep{ComplexAnalysis}.
More recently there has been a move to approximate Cauchy transforms (and more generally singular integrals) via a spectral method, as outlined by \citet{slevinsky2017fast, olver2011computing}, and implemented numerically in the Julia software packages \texttt{SingularIntegralEquations}, \texttt{ApproxFun} and \texttt{WienerHopf}. An outline of this method is to expand a function in terms of weighted orthogonal polynomials and then compute the Cauchy transform on those polynomials to give a highly accurate numerical method.

\Citet{trogdon2015riemann} and \citet{Elena} show that various mappings on the interval, $\mathbb{I}$, can be used in conjunction with Plemelj's lemma to prove that a Cauchy transform of a contour may be expressed as mapped Cauchy transforms. The implementation here makes use of two mappings in particular, an affine map $p(k) = a + b k $, and the real line map developed by \citet{Elena}, $r(k) = \frac{ k + k^3 }{(1-k^2)^2}$. Observe that by using both mappings together, one can map the contours $\Gamma_{\pm}$ to the interval $\mathbb{I}$, this can be done in the following way $\Gamma_{\pm} = p(r([-1,1])) = \{ p(r(x)) : -1 \leq x \leq 1 \}$ thus
\begin{align}
  \mathcal{C}_{\Gamma_{\pm}}[f](k) =&  \mathcal{C}_{\mathbb{R}} [f \circ p ] (p^{-1}(k)) \\
  =& {\sum_{j = 1}}^4 \mathcal{C}_{[-1,1]} [f \circ p \circ r] ({r_j}^{-1}(p^{-1}(k))) \\
  &- 2\mathcal{C}_{[-1,1]} [f \circ p \circ r] (p^{-1}(1)) -  22\mathcal{C}_{[-1,1]} [f \circ p \circ r] (p^{-1}(-1)) .
\end{align}
Where the functions $p^{-1} $ and $ {r_j}^{-1} $ are inverses to $p$ and $r $ respectively, and the subtracted terms are to remove the behaviour at infinity. The Cauchy transform of the $\Gamma_{\pm}$ contours can be expressed as the composition of multiple mappings as each mapped Cauchy transform satisfies Plemelj's lemma.

Finally, there is some special consideration required when decomposing the oscillatory forcing terms, as the oscillatory behaviour leads to numerical instabilities and exponential growth. To combat these issues, one may deform the integration contour onto the steepest descent contour for the integrand, which turns the oscillatory behaviour into exponential decay. In calculating the Cauchy transforms over the steepest descent contours, quadratic maps have been used to represent the contour. There is an alternative developing method by \citet{trogdon2016}, which uses a special polynomial basis for evaluating the Cauchy transform of oscillatory functions, however this work is ongoing.

\section{Free-boundary problem}\label{sec:Free-boundary}

Throughout it has been assumed that the unknown junctions points of the boundary conditions, $-a$ and $d$, are known. Taking this assumption allows a solution to be found where the junction points may be viewed as parameters to the system. The assumption enables the application of the Wiener-Hopf technique to find a solution and then we implement an optimisation method which finds the correct junction points a posteriori. Validation of the junction points is found by ensuring continuity of the solution as it approaches $-a$ and $d$. The secant method is implemented here to re-estimate the junction points by finding the roots of the solution.

We recall
 the edge conditions
\begin{align}
    {\tau_{yy}}|_{y=0}(x) \rightarrow 0 &\text{ as } x \rightarrow -a^{+} ,  
    &&{\tau_{yy}}|_{y=0}(x) \rightarrow 0  \text{ as } x \rightarrow d^{-}.
\end{align}
Imposing the edge conditions gives a route for imposing an optimisation method. To formulate the optimisation problem more formally,
the free-boundary problem is cast as the following optimisation problem
\begin{align} \label{eqn: FSMin}
 \min_{\chi^j} \| &{\tau_{yy}^j}( \chi^j , 0 )  \|.
\end{align}
Where we define the initial and $j$-th iteration junction points and solution which corresponds as 
\begin{align}
    \chi^0 &= (a^0 , d^0 ) , &\chi^j = (a^j , d^j ), && \tau^{0}(x) = {\tau_{yy}}^0( x , 0 ),
     && \tau^j(x) = {\tau_{yy}}^j( x , 0 ). 
\end{align}
To find an initial solution, we take two initial guesses of the junction points, $\chi^0, \chi^1$ and find the corresponding solutions ${\tau_{yy}}^0( x , 0 ), {\tau_{yy}}^1( x , 0 )$ . Then we find the next iteration of junction point by implementing a secant method. We may find the $j+1$-th iteration of junction points, $\chi^{j+1} $, by solving the following equations
\begin{equation}
    \chi^{j+1} = \left(\! a^{j} - {\tau}^j\big({-}a^{j}\big)\!\times\! \left( \frac{a^j - a^{j-1}}{\tau^j(-a^j) - \tau^{j-1}( -a^{j-1})} \right)\!,\; d^{j} - {\tau}^j\big( d^{j} \big)\!\times\! \left( \frac{d^j - d^{j-1}}{\tau^j (d^j) - \tau^{j-1}(d^{j-1})} \right)\!\right)\!.
\end{equation}
 This procedure is iterated until the evaluation is below a tolerance,
$$
\| \tau^j ( \chi^{j} ) \| \leq \text{tol} .
$$
Once converged the final $\chi^j$ will give a solution where the junction points are accurately estimated and ensure continuity of the solution. 

\begin{figure}
    \centering
    \includesvg[width=0.7\textwidth]{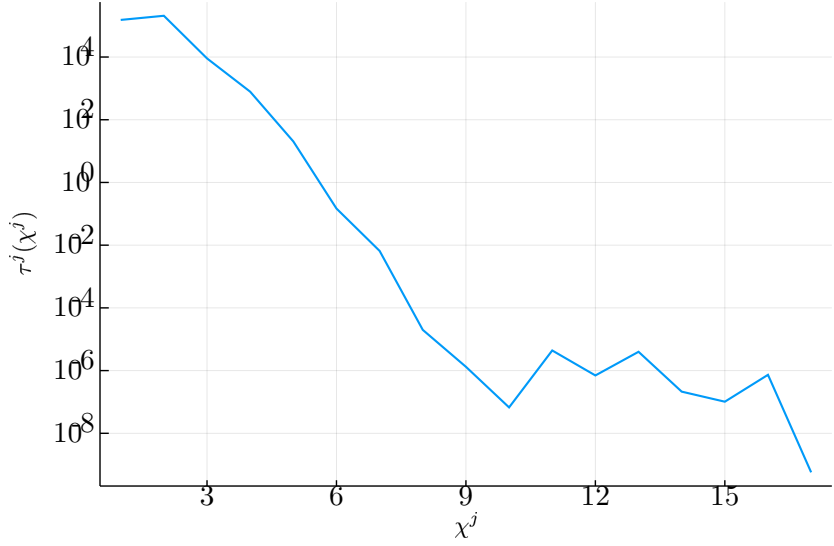} % second figure i
    \caption{ This figure shows the evaluation for $\tau_{yy}(\chi^j,0)$ for differing $-a$ and $d$. The optimisation method is implemented, showing that as iterations further, the solution converges. The method is terminated when $\tau_{yy}(\chi^j,0)$ is smaller than the tolerance, which is found after 17 iterations in this case.} \label{fig: FSFree}
\end{figure}

The figure \ref{fig: FSFree} shows the profile of $\tau_{yy}(x,0)$ for the first 17 iterations of the free-boundary method. The method terminates once the evaluation of $\tau_{yy}$ is below the preset tolerance of $1e-8$, which requires many iterations but is quite close to the converged solution after the third iteration. It is clear from the figure that the converged iterations have a smooth continuous solution, as one would expect.

\section{Results}\label{sec:Results}

The application of the iterative procedure gives an approximation to the terms $ \wdft{v}{d}{}{+}, \wdft{v}{-a}{}{-}, \wft{\tau}{d}{}{-}$ and $\wft{\tau}{-a}{}{+} $, from which one can numerically evaluate the unknown functions $\hat{A_1}, A_2$, and therefore the solution $\mathbf{u}(x,y)$ . To find the solution and stresses the Fourier transforms must be inverted, which is computed via a Gaussian quadrature rule. To illustrate, the stress may be evaluated as follows, 
\begin{align} 
  \tau_{yy}(x,0) =& \frac{1}{{2\pi}}\int^{\infty}_{-\infty} \widetilde{\tau_{yy}}(k,0) e^{-ikx} dk \\
                   =& \frac{1}{{2\pi}}\int^{\infty}_{-\infty}  \wft{{\tau_{yy}}_{y=0}}{d}{}{-}(k) e^{ik(d-x)} dk \\
                   \approx & \sum_{j=0}^{n} w_j{\wft{\psi}{d}{}{-}(k_j)} e^{ik_j(d-x)} .
\end{align}
Where $k_j$ are the nodes and $w_j$ are the weights for the quadrature rule.
Then the approximation for $\tau_{yy}(x,0)$ may be found by closing the contour in either the upper or lower half-plane for $\wft{{\tau_{yy}}_{y=0}}{-a}{}{+}(k)$ or $\wft{{\tau_{yy}}_{y=0}}{d}{}{-}(k) $ respectively, giving
\begin{align} \label{eqn: FSsol tau}
  \tau_{yy}(x,0) = 
  \begin{cases}
  0,&   \quad  d < x ,  \\
  \frac{1}{{2\pi}}\int^{\infty}_{-\infty} \widetilde{\tau_{yy}}(k,0) e^{-ikx} dk,&  -a \leq x \leq d, \\
  0,& \quad  x < -a.
  \end{cases}
\end{align}
A plot of the solution of the normal stress, $\tau_{yy}(x,0)$, for given parameter values is shown in figure \ref{fig: FSnormal stress}. The plot validates the method used somewhat as it follows the behaviour one would expect from the boundary conditions, namely that the normal stress on the free-boundary is equal to zero.

Again by contour integration one can deduce that
\begin{align}
v(x,0) = 
\begin{cases}
\frac{1}{{2\pi}}\int^{\infty}_{-\infty}  \wdft{v}{d}{}{+}(k) e^{ik(d-x)}  dk,&   \quad  d < x ,  \\
\frac{1}{{2\pi}}\int^{\infty}_{-\infty}  f^d_{-a}(k)e^{-ikx} dk,&  -a \leq x \leq d, \\
\frac{1}{{2\pi}}\int^{\infty}_{-\infty}  \wdft{v}{-a}{}{-}(k) e^{-ik(a+x)}  dk,& \quad  x < -a .
\end{cases}
\end{align}
An approximation to the terms $\tau_{yy}(x,0)$ and $ v(x,0) $ has been made, so by the using of \eqref{eqn: FSty} and \eqref{eqn: FSv} an approximation to the displacement terms may be made.

The figures in \ref{fig: FSExample} illustrate the solution with the parameters based on that of structural steel~\citep{Eurocode}. The continuity at the junction points for $\tau_{yy}( x ,0)$ is ensured by applying the optimisation method detailed in the proceeding section. A contour plot of the distribution of the stresses within the elastic material is included, which shows the concentration of the stresses to be around the cylinder and away from the surface. The stresses are calculated by evaluating \eqref{eqn: FSregularised solution} through a Gaussian quadrature rule and relating it to $\tau_{xy}(x,y)$ and $\tau_{yy}(x,y)$. The traction on the surface of the elastic media may be found to be $\vect{T} = (-\tau_{xy}(x,0) ,- \tau_{yy}(x,0) )$, which in this case faces in the positive-$x$ and positive-$y$ direction.

\begin{figure}
    \centering
        \includesvg[pretex=\scriptsize,width=0.5\textwidth]{Another} % second figure itself    
        \caption{This figure shows the plot $\tau_{yy}(x,0)$ profile under the parameter values below.}
  \label{fig: FSnormal stress}\par\medskip
    \begin{minipage}{0.48\textwidth}
        \centering
        \includesvg[pretex=\tiny,width=\textwidth]{NormalStress_med} % first figure itself         
        \caption{Here we see the stress distribution for, $\tau_{yy}(x,y)$, under the same parameter values below.  }
    \label{fig: FSnormal}
    \end{minipage}\hfill
   \begin{minipage}{0.48\textwidth}
        \centering
    \includesvg[pretex=\tiny,width=\textwidth]{ShearStress_med} % second figure itself        
  \caption{Here we see the stress distribution for, $\tau_{xy}(x,y)$, under the same parameter values below.}
  \label{fig: FSshear 1}
    \end{minipage}
    \caption{These three figures show plots of the stress distribution for the parameter values $\lambda = 210000, \mu =  81000, \mu_0 = 0.3, \rho = 7850, R = 100, V = 1$ and $\omega = -0.5i$. The contact region was found to be $ -0.7763734861967965, 0.9220015375514687 $. These figures validate the method used as we have the continuity of $\tau_{yy}(x,0)$ and the behaviour of in the elastic media is as one would expect. }
    \label{fig: FSExample}
\end{figure}

\subsection{Parameter study}

The technique implemented here is valid for any choice for the parameters $\lambda, \mu, \mu_0$ and $ \rho$ provided the speed of the cylinder, $V$, is adjusted to ensure that a strip of analyticity encapsulating the real line remains. The requirement on $V$ is then 
\begin{align}
    V^2 < {\frac{\mu}{\rho}},
\end{align}
which will lead to singular points and branch cuts of $C_{1}(k)$ away from the real line. However, the closer $V$ is to $0$ leads to solutions which are simpler to decompose as the singular points or branch cuts are more equidistant from the real line. To illustrate a solution for different parameters, the figures \ref{fig: FSnonmetal normal}, \ref{fig: FSnonmetal displacement} ,and \ref{fig: FSnonmetal stress} show plots of the profile of $\tau_{yy}(x,0)$, magnitude of displacement $|\vect{u}(x,y)|$, and distribution of $\tau_{yy}(x,y)$ respectively. The figure \ref{fig: FSnonmetal normal} shows that there exists a unique solution to the free-boundary problem for the chosen parameters. As there is little dependency on the material parameters we shall include plots with the parameters set to be similar to structural steel.

\begin{figure}
    \centering
    \includesvg[pretex=\tiny,width=0.5\textwidth]{V_y0} % second figure itself
        \caption{This figure shows the plot $v(x,0)$ profile under the parameter values below.}
  \label{fig: FSnonmetal normal}\par\medskip
    \begin{minipage}{0.48\textwidth}
        \centering
        \includesvg[pretex=\tiny,width=\textwidth]{ShearStress} % first figure itself
        \caption{This figure shows the plot $\tau_{xy}(x,y)$ profile under the parameter values below.  }
    \label{fig: FSnonmetal displacement}
    \end{minipage}\hfill
   \begin{minipage}{0.48\textwidth}
        \centering
    \includesvg[pretex= \tiny,width=\textwidth]{NormalStress} % second figure itself
  \caption{This figure shows the plot $\tau_{yy}(x,y)$ profile under the parameter values below.}
  \label{fig: FSnonmetal stress}
    \end{minipage}
    \caption{These two figures show plots of the solution for the parameter values $\lambda = 10000, \mu =  1000000, \mu_0 = 0.3, \rho = 1250, V = 10$ and $\omega = -0.5i$. Here the contact region was taken to be $-0.8837359101451169, 0.8792489231188044$. These figures validate the method the continuity of $\tau_{yy}(x,0)$ and the behaviour of $\vect{u}(x,y)$ is as one would expect.}
\end{figure}

\subsubsection{Forward and backwards slip}

In the nonlinear derivation of the boundary conditions it was found that either forward or backwards slip were valid boundary conditions, with each representing a different physical system, either wheel spinning or locking respectively. To illustrate that these two physical systems are indeed realisable, the figures \ref{fig: FSForward} and \ref{fig: FSBackward} have been included, showing the existence of a solution for both slipping configuration under the same parameters.

The figures \ref{fig: FSForward} and \ref{fig: FSBackward} show a comparison of the solution found here for both forward slip and backwards slip boundary conditions, with the same parameters are assumed in each. The junction points $-a$ and $d$ for both slipping configurations are found to be different, suggesting the direction of slip being an important factor in determining the location of junction points. A notable difference between the two slipping configurations may be seen in the figures \ref{fig: FSSW FS displacement} and \ref{fig: FSSW BS displacement}, which shows the distribution of $\tau_{yy}(x,y)$ for both configurations and in particular that the direction where the elastic media experiences the most stress differs. The difference of forward and backward slip is verified in the plots \ref{fig: FSSW FS normal stress} and \ref{fig: FSSW BS normal stress}, which shows opposing signs for $\tau_{xy}(x,0)$. The physical difference of these two slip directions may be seen in the traction too, as traction for forward slip faces in the positive-$x$ and positive-$y$ direction whereas traction for backward slip faces in the negative-$x$ and positive-$y$ direction.

\begin{figure}
    \centering
    \textbf{Forward Slip}\par\medskip
    \begin{minipage}{0.48\textwidth}
        \centering
        \includesvg[pretex=\tiny,width=\textwidth]{Tau_xy} % second figure itself
        \caption{This figure shows the plot $\tau_{xy}(x,0)$ profile under the parameter values below.}
  \label{fig: FSSW FS normal stress}
    \end{minipage}\hfill
    \begin{minipage}{0.48\textwidth}
        \centering
        \includesvg[pretex=\tiny,width=\textwidth]{NormalStress_FS} % first figure itself
        \caption{Here we see the stress distribution for, $\tau_{yy}(x,y)$, contour under the same parameter values below.  }
    \label{fig: FSSW FS displacement}
    \end{minipage}
    \caption{These two figures show plots of the solution for the parameter values $\lambda = 210000, \mu =  81000, \mu_0 = 0.3, \rho = 7850, V = 1$ and $\omega = -0.5i$. Here the contact region was taken to be $-0.7763734861967965, 0.9220015375514687$.}
    \label{fig: FSForward}
\end{figure}

\begin{figure}
    \centering
    \textbf{Backward Slip}\par\medskip
    \begin{minipage}{0.48\textwidth}
        \centering
        \includesvg[pretex=\tiny,width=\textwidth]{Tau_xy_y0} % second figure itself
        \caption{This figure shows the plot $\tau_{xy}(x,0)$ profile under the parameter values below.}
  \label{fig: FSSW BS normal stress}
    \end{minipage}\hfill
    \begin{minipage}{0.48\textwidth}
        \centering
        \includesvg[pretex=\tiny,width=\textwidth]{NormalStress_BS} % first figure itself
        \caption{Here we see the stress distribution for, $\tau_{yy}(x,y)$, contour under the same parameter values below.  }
    \label{fig: FSSW BS displacement}
    \end{minipage}
    \caption{These two figures show plots of the solution for the parameter values $\lambda = 210000, \mu =  81000, \mu_0 = 0.3, \rho = 7850, V = 1$ and $\omega = -0.5i$. Here the contact region was taken to be $-0.8251365655140489, 0.8574926897492611$.}
    \label{fig: FSBackward}
\end{figure}

\subsubsection{Small frequency limit}

In the approach here we introduced the temporal frequency, $\omega$, into the governing equations to generate a strip of regularity in the complex $k$-plane. However, physically we are interested in the steady state case which corresponds to the limit as $\omega \rightarrow 0$. Here we shall discuss the effects of taking the limit of $\omega \rightarrow 0$ and seek to address the following:
\begin{enumerate}
    \item Does the limit $\omega \rightarrow 0$ converge?
    \item What is a good approximation to $\omega = 0 $?
\end{enumerate} 

\begin{figure}
    \centering
    \includesvg[width=0.7\textwidth]{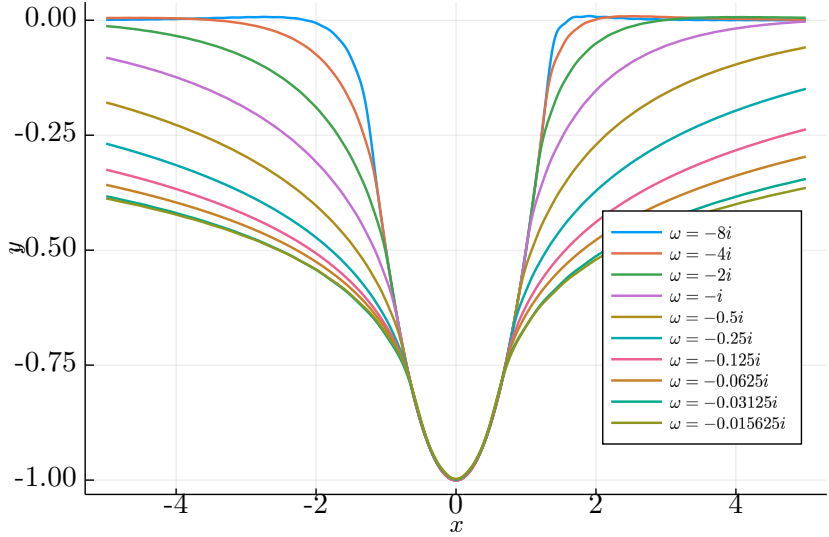}    
    \caption{Plots of $v(x,0)$ for the following parameter values $\lambda = 210000, \mu =  81000, \mu_0 = 0.3, \rho = 7850 $ and $V = 1$, with varying $\omega$ and forward slip.}
    \label{fig: FSvomega}
\end{figure}

The introduction of $\omega$ is to allow a strip of regularity for the Wiener-Hopf technique to be applied. The quantity which this impacts the most is the function $K(k)$, which is multiplicatively decomposed. Without the introduction of $\omega$, $K (k)$ would have two branch cuts extending from the origin to infinity, but by introducing an imaginary $\omega$ the branch points are separated to above and below the origin. In some sense the temporal frequency parameterises the branch points and hence the strip. Taking the limit of $\omega \rightarrow 0$ coalesces these branch points and the solution converges as seen in figure~\ref{fig: FSvomega}.

\begin{figure}
        \centering
        \includegraphics[width=0.7\textwidth]{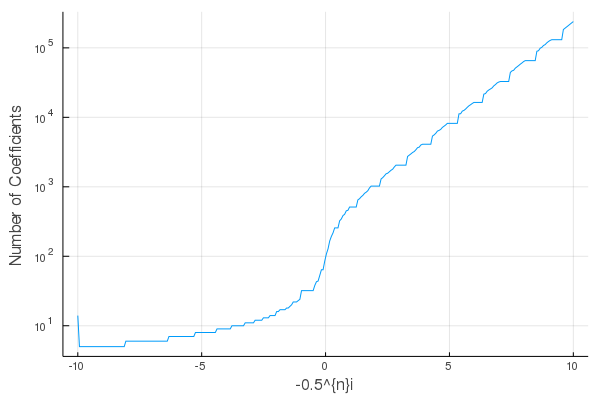} % first figure itself
        \caption{The number of coefficients needed to approximate $K(k,\omega)$ for decreasing $\omega$.  These two figures show plots of the error of introducing $K(k,\omega)$ in approximating $K(k, 0 )$ for the parameter values $\lambda = 210000, \mu =  81000, \mu_0 = 0.3, \rho = 7850 $ and $V = 1$ .}
        \label{fig:Coeffs2}
\end{figure}

To address the first question, the figure \ref{fig:Coeffs2} shows that decreasing $\omega$ does decrease the error and $K$ converges to the $\omega = 0$ case. However, as seen there is a blowup near the origin, where $K(k,0)$ approaches $\frac{0^4}{0^3}$, which is actually $0$ but computationally returns a singularity. We may avoid the singularity by setting $K(0,0)$ to be $0$, which leads to another singularity when taking the multiplicative decomposition as we have $log(0)$. Another question is whether different ways the branch cuts coalesce leads to different solutions, the figure \ref{fig:Coeffs2} addresses this by showing that $C_{1}(k)$ regardless of the argument of $\omega =  re^{i\theta}$.
The singularity in the multiplicative decomposition increases the computational cost of taking the Cauchy transform, as it is required that the function be Lipschitz continuous to approximate it in an orthogonal polynomial basis. We may still find an expansion but as we decrease $\omega$ we increase the number of terms needed for the expansion, see fig \ref{fig:Coeffs2}. The computational cost of finding the expansion increases with the number of terms linearly, but then calculating the Cauchy transform becomes very computationally costly. There are two solutions we may use for this, approximate $\omega = 0$ where we may take $\omega =  - 0.01i $ as a good approximation, or eliminate the terms in the expansion which are below a certain threshold which still means calculating the expansion is costly but subsequent Cauchy transforms are less so.

In summary, we see that the solution converges in the limit $\omega \rightarrow 0$ and that it is unique regardless of the ways the branch cuts may coalesce. In the case $\omega = 0$ we are faced with increased computational costs in taking the multiplicative decomposition but approximating $\omega = -0.5 i$ works well without being too costly.

To observe how the solution behaves as $\omega$ varies, the figures \ref{fig: FStauomega} and \ref{fig: FSvomega} have been included. Due to the presence of $\gamma_1(k)$ and $\gamma_2(k)$ in the solution, one would expect the solution to decay faster for larger $\omega$, which is seen from the two sets of figures. For small $\omega$, there appears to be very little difference between the plots of the vertical profile and the stress distribution, suggesting that $\omega = -0.5 i$ is a sufficiently small approximation.

\begin{figure}
    \centering
    \begin{minipage}{0.48\textwidth}
        \centering
        \includesvg[pretex=\tiny, width=\textwidth]{NormalStress25}
        \caption{$\omega =  -0.25i$. }
    \label{fig: FSt0.25}
    \end{minipage}\hfill
    \begin{minipage}{0.48\textwidth}
        \centering
        \includesvg[pretex=\tiny,width=\textwidth]{NormalStress2}
        \caption{$\omega =  -2i$. }
    \label{fig: FSt2}
    \end{minipage}\vfill
    \begin{minipage}{0.48\textwidth}
        \centering
        \includesvg[pretex=\tiny,width=\textwidth]{NormalStress5}
        \caption{$\omega =  -0.5i$. }
    \label{fig: FSt0.5}
    \end{minipage}\hfill
    \begin{minipage}{0.48\textwidth}
        \centering
        \includesvg[pretex=\tiny,width=\textwidth]{NormalStress4} % first figure itself
        \caption{$\omega =  -4i$. }
    \label{fig: FSt4}
    \end{minipage}
    \vfill
    \begin{minipage}{0.48\textwidth}
        \centering
        \includesvg[pretex=\tiny,width=\textwidth]{NormalStress1} % first figure itself
        \caption{$\omega =  -i$. }
    \label{fig: FSt1}
    \end{minipage}\hfill
    \begin{minipage}{0.48\textwidth}
        \centering
        \includesvg[pretex=\tiny,width=\textwidth]{NormalStress8} % first figure itself
        \caption{$\omega =  -8i$. }
    \label{fig: FSt8}
    \end{minipage}
    \caption{Plots of $\tau_{yy}(x,y)$ for the following parameter values $\lambda = 210000, \mu =  81000, \mu_0 = 0.3, \rho = 7850 $ and $V = 1$, with varying $\omega$.}
    \label{fig: FStauomega}
\end{figure}

\subsection{Von Mises yield criterion}

To reintroduce elastoplastic modelling into the problem, we may consider the locations where the media begins to transition to elastoplastic behaviour. The location where the media yields may be found by the following equation,
\begin{equation}
   ( \tau_{xx} - T ) ^2 + 2\tau_{xy}^2 +  (\tau_{yy} - T )^2 \leq \tau_{Y}^2,
\end{equation}
where $T = (\tau_{xx} + \tau_{yy} + \tau_{zz})/3 $ which may be interpreted as the location where the elastic energy reaches a critical value.

The figures \ref{fig: FSvon BS displacement} and \ref{fig: FSvon BS normal stress} shows the magnitude of yield in the elastic media for both forwards and backwards slip. In the regions directly beneath the cylinder, the magnitude of elastic energy is greatest for both, suggesting that these are the regions where the plastic behaviour is likely to occur.

\begin{figure}
    \centering
    \begin{minipage}{0.48\textwidth}
        \centering
    \textbf{Forward Slip}\par\medskip
        \includesvg[pretex=\tiny,width=\textwidth]{VonMises_FS} % second figure itself
        \caption{This figure shows the magnitude of the Von Mises criterion under the parameter values below.}
  \label{fig: FSvon BS normal stress}
    \end{minipage}\hfill
    \begin{minipage}{0.48\textwidth}
        \centering
    \textbf{Backward Slip}\par\medskip
        \includesvg[pretex=\tiny,width=\textwidth]{VonMises_BS} % first figure itself
        \caption{This figure shows the magnitude of the Von Mises criterion under the parameter values below.}
    \label{fig: FSvon BS displacement}
    \end{minipage}
    \caption{These two figures show plots of the solution for the parameter values $\lambda = 210000, \mu =  81000, \mu_0 = 0.3, \rho = 7850, V = 1$ and $\omega = -0.5i$. The contact region is for each configuration is the same as in  figures \ref{fig: FSForward} and \ref{fig: FSBackward}.}
\end{figure}

\section{Conclusion}\label{sec:Conclusion}

We have developed a model of an elastic half-space deformed by a cylindrical roller. The friction law assumed here is that of full-slip, which leads to a $2 \times 2$ matrix Wiener--Hopf problem. Due to the positions of the exponential terms within the matrices an iterative method which initially decouples the scalar Wiener--Hopf equations is suitable. Finally, the free-boundary problem is solved a posteriori by making an initial estimate and iteratively finding better estimates from there.

The method employed here requires the introduction of a small wavenumber to introduce a strip of analyticity, which enables a matrix Wiener--Hopf problem to be constructed. This may be bypassed by framing the problem as a Riemann-Hilbert problem and solving that numerically~\citep{trogdon2015riemann}. Alternative approaches may be suitable for the free-boundary problem too, with it baring great similarities to floating body problems.

The limit of full-slip is interesting due to the simplification that it provides to the matrix Wiener--Hopf over the frictional case. The method employed here applies to a stick-slip friction law but the matrix Wiener--Hopf will be considerably more challenging to solve. The full-stick limit provides a scenario where a special case adaptation of the iterative method~\citep{Kisil2018} may be required, as the scalar Wiener--Hopf equations cannot be decoupled easily.

\subsection*{Acknowledgements}

HB was supported in this work through the University of Warwick MASDOC Doctoral Training Centre, and gratefully acknowledges their support.
EJB gratefully acknowledges the support of a Royal Society University Research Fellowship (UF150695 and RGF\textbackslash EA\textbackslash 180284), and of a UK Research and Innovation (UKRI) Future Leaders Fellowship (MR/V02261X/1).

\appendix
\section{Derivation of the governing equations}
\label{app:governing-equations}

\begin{figure}%
\centering%
\includegraphics{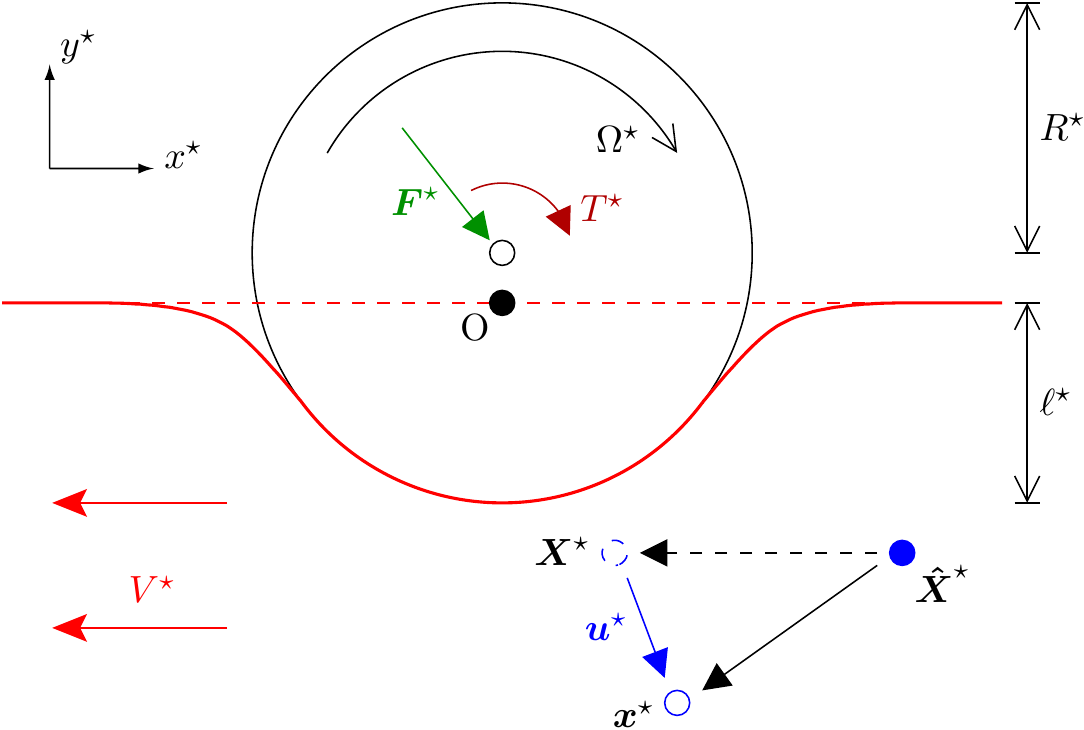}%
\caption{Schematic of a cylinder rolling along an elastic half-space.  The cylinder moves at a linear velocity $V^{\star}$ in the $x^{\star}$-direction along the surface of the elastic half-space.  The origin of the coordinate system (labelled~O) is taken in a frame of reference moving with the cylinder, directly below the centre of the cylinder at the height of the undeformed elastic surface.  The cylinder of radius $R^\star$ rolls about its centre axis with angular velocity $\Omega^\star$, and a force $\vect{F}^{\star}$ and torque $T^{\star}$ are applied to the centre of the cylinder, causing the cylinder to be indented by $\ell^\star$ into the elastic half-space.  A material point labelled by its location $\vect{\hat{X}}^{\star}$ at time $t^{\star}=0$ would have moved to location $\vect{X}^{\star}$ at time $t^{\star}$ without the cylinder being present, but has instead moved to a location $\vect{x}^{\star}$ due to the deformation caused by the cylinder, giving an elastic displacement $\vect{u}^{\star} = \vect{X}^{\star} - \vect{x}^{\star}$.}
\label{fig: FSschematic}
\end{figure}

We consider the situation shown schematically in figure~\ref{fig: FSschematic}.
In what follows, a star denotes a dimensional quantity.
 
 A cylinder of radius $R^\star$ is pushed into an elastic half-space $y^{\star} < 0$ with a force $\vect{F}^{\star}$, resulting in an indentation of depth $\ell^\star$.  The cylinder is rotated with a torque $T^{\star}$, giving an angular velocity $\Omega^\star$, and the cylinder therefore translates in the $x^{\star}$-direction across the half-space at a linear velocity $V^{\star}$.  We choose a frame of reference moving with the cylinder, such that the centre of the cylinder is located at $x^{\star}=0$ and $y^{\star}=R^\star-\ell^\star$, with $\vect{e_x}$ and $\vect{e_y}$ unit vectors in the $x^{\star}$- and $y^{\star}$-directions respectively.  We label material particles in the elastic half-space by the location $\vect{\hat{X}}^{\star}$ at time $t^{\star}=0$ in the absence of the cylinder.  In the absence of the cylinder, such a material particle would be located at $\vect{X}^{\star} = \vect{\hat{X}}^{\star} - V^{\star}t^{\star}\vect{e_x}$ at time $t^{\star}$.  With the cylinder present, the material particle has been displaced, and is instead located at $\vect{x}^{\star} = \vect{X}^{\star} + \vect{u}^{\star}$, where $\vect{u}^{\star} = (u^{\star},v^{\star})$ is the displacement in the elastic half-space.

\subsection{Nonlinear dimensional governing equations}

We derive the following governing equation in the Eulerian configuration. In the Eulerian configuration, a material particle is referred to using its deformed location $\vect{x}^{\star}$, so the displacement $\vect{u}^{\star}(\vect{x}^{\star},t^{\star})$ and the undeformed position $\vect{X}^{\star}(\vect{x}^{\star},t^{\star})$ are both functions of $\vect{x}^{\star}$ and $t^{\star}$. Thus the displacement in the Eulerian configuration is defined to be
\begin{equation}
    \vect{u}^{\star}(\vect{x}^{\star},t^{\star}) = \vect{x}^{\star} - \vect{X}^{\star}(\vect{x}^{\star},t^{\star}).
\end{equation}
To calculate the velocity of a material particle, let $\vect{\hat{x}}^{\star}(t^{\star},\vect{\hat{X}}^{\star})$ be the location of a material particle at time $t^{\star}$ whose undeformed location was $\vect{\hat{X}}^{\star}$ at time $t^{\star}=0$.  Thus,
\begin{equation}
    \vect{\hat{x}}^{\star}(t^{\star}) = \vect{X}^{\star}(\vect{\hat{x}}^{\star}(t^{\star}),t^{\star}) + \vect{u}^{\star}(\vect{\hat{x}}^{\star}(t^{\star}),t^{\star})
    = \vect{\hat{X}}^{\star} - V^{\star}t^{\star}\vect{e_x} + \vect{u}^{\star}(\vect{\hat{x}}^{\star}(t^{\star}),t^{\star}).
\end{equation}
Taking the time derivative with $\vect{\hat{X}}^{\star}$ fixed, we find
\begin{equation}
\vect{v}^{\star} = \frac{\partial\vect{\hat{x}}^{\star}}{\partial t^{\star}}\Bigr|_{\vect{\hat{X}}^{\star}} =  - V^{\star}\vect{e_x} + \frac{\partial \vect{u}^{\star}}{\partial t^{\star}}\Bigr|_{\vect{x}^{\star}} + \vect{v}^{\star}\cdot\frac{\partial \vect{u}^{\star}}{\partial \vect{x}^{\star}}\Bigr|_{t^{\star}}\label{equ:eulerian-velocity}.
\end{equation}
which is an implicit equation for the material particle velocity $\vect{v}^{\star}(\vect{x}^{\star},t^{\star})$ (not to be confused with the vertical component of the displacement vector $v^\star$).

Newton's law of motion, or equivalently conservation of momentum, gives the governing equations in the bulk of the material as
\begin{equation}\label{equ:eulerian-momentum}
\rho^\star\!\left(\frac{\partial\vect{v}^{\star}}{\partial t^{\star}} + \vect{v^{\star}\cdot}\frac{\partial\vect{v}^{\star}}{\partial\vect{x}^{\star}}\right) = \vect{\nabla^{\star}\cdot}\mat{\tau^{\star}},
\end{equation}
{where $\mat{\tau}^{\star}$ is the Cauchy stress tensor.}

\subsection{Nonlinear dimensional boundary conditions}

We define the surface of the elastic half-space to be $y^{\star} = \eta^{\star}(x^{\star})$.  Where the elastic half-space is in contact with the cylinder, for $-a^{\star} \leq x^{\star} \leq d^{\star}$, the elastic displacement is constrained to not penetrate the cylinder, and the material slips~\citep{Barber_contact} according to Coulomb friction.  Outside this contact region, the surface is stress free.

\subsubsection{No penetration boundary condition}\label{sec:contact}

The no penetration boundary condition comes from the physical condition that the elastic half-space cannot penetrate the rigid cylinder.   This means that a material particle on the surface of the elastic half-space (given by $Y^{\star}=0$), when in contact with the cylinder, must lie on the surface of the cylinder a distance $R^\star$ from the cylinder center.  By writing $y^{\star} = Y^{\star} + v^{\star}$, and noting that the cylinder centre is located at $(0,R^\star-\ell^\star)$, we deduce that 
\begin{align}
{x^{\star}}^2 + \big(y^{\star} -R^\star+\ell^\star\big)^2 &= {R^\star}^2 &
&\text{when}&
-a^{\star} < x^{\star} < d^{\star}
\qquad&\text{and}\qquad
Y^{\star} = 0.
\label{equ:lagrangian-contact}
\end{align}

\subsubsection{Slipping boundary condition}

The slipping boundary condition comes from the physical condition that the elastic media is slipping past the rigid cylinder, so follows the Coulomb law of dry friction, $\vect{n^\star\cdot}\mat{\tau}^{\star}\vect{\cdot t^{\star}} = \pm \mu_{0} (\vect{n^\star\cdot}\mat{\tau}^{\star}\vect{\cdot n^\star})$, where $\mat{\tau^\star}$ is the Cauchy stress tensor.  The (non-unit) normal to the cylinder may be taken as $\vect{n}^{\star} = (x^{\star}, y^{\star}-R^\star+\ell^\star)$, and the tangent vector to the cylinder may be taken as $\vect{t}^{\star} = ( y^{\star}-R^\star+\ell^\star, -x^{\star})$.  This results in the Coulomb friction law
\begin{multline}\label{eqn: FSnonlinslip}
  \frac{x^{\star}}{(y^{\star} - R^\star + \ell^\star) }(\tau_{x x}^{\star} {-} \tau_{y y}^{\star}) + \left(1 {-} \left(\frac{ x^{\star}}{y^{\star} - R^\star + \ell^\star}\right)^{\!2}\right) \tau_{x y}^{\star}  \\
  =\pm\mu_{0}\left[ \left(\frac{ x^{\star} }{y^{\star} - R^\star + \ell^\star}\right)^{\!2}\!\tau_{ x x }^{\star}  +  \frac{2x^{\star}}{y^{\star} - R^\star + \ell^\star}\tau_{x y}^{\star}  + \tau_{y y}^{\star} \right]. 
\end{multline}
With this convention for the directions of the normal and tangent vectors, $\pm = +$ if the cylinder is exerting a friction force on the elastic half-space in the negative $x^{\star}$-direction, meaning that the angular velocity $\Omega^\star$ is sufficiently large and positive (a wheel-spin type condition), where as $\pm = -$ if $\Omega^\star$ is sufficiently negative (a locked-wheel braking type condition).  Alternatively, to determine the direction of slip, the velocity of the surface of the cylinder in the clockwise direction, $v_c^{\star} = R^\star\Omega^\star$, may be compared to the velocity of the elastic material in the same direction, $v_e^{\star} = \vect{v^{\star} \cdot t^{\star}}/|\vect{t^{\star}}|$: if $v_c^{\star} > v_e^{\star}$ then $\pm=+$; while if $v_c^{\star} < v_e^{\star}$ then $\pm=-$.  Note that $v_e^{\star}$ is a function of the location $x^{\star}$, and that in this paper we assume that one of these inequalities holds throughout the contact region, so that the entire contact surface is slipping; in other words, there are no regions of sticking, only slip.

\subsubsection{Stress-free boundary condition}

The final boundary condition is that outside of the contact region, $-\infty < x^{\star} < -a^{\star} $  and $ d^{\star} < x^{\star} < \infty $, the surface is free of traction, and so $\mat{\tau}^{\star}\cdot \vect{n^\star} = \vect{0}$. Here, the (non-unit) normal to the deformed free surface $y^{\star} = \eta^\star(x^{\star})$ is given by $\vect{n^\star} = ( -\intd \eta^\star/\intd x^{\star},\, 1  )$.  This gives the boundary conditions
\begin{align}\label{eqn: FSfree}
-\frac{\intd \eta^\star}{\intd x^{\star}} \tau_{x x}^{\star} +  \tau_{x y}^{\star} = -\frac{\intd \eta^\star}{\intd x^{\star}} \tau_{ xy}^{\star} +  \tau_{y y}^{\star} &=  0 &
&\text{when}\quad
y^{\star} = \eta^\star(x^{\star}).
\end{align}

\subsection{Nondimensionalization}

In what follows, we will derive linear governing equations and boundary conditions.  Linearization can only be justified provided there is a small parameter, which in this case is due to the indentation into the elastic half space $\ell^\star$ being much smaller than the cylinder radius $R^\star$.  We first make this explicit by nondimensionalizing the situation described above.  Dimensionless variables are denoted without a star.

We first note that the indentation $\ell^\star$ is expected to induce an elastic displacement of the same order, so that $\vect{u^\star} = O(\ell^\star)$.  However, the contact region is given by the length of the cylinder in contact with the elastic half space, which is bounded by $\sqrt{{R^\star}^2 - (R^\star - \ell^\star)^2} = \sqrt{2R^\star\ell^\star}(1 + \ell^\star\!/2R^\star)^{1/2}$.  We therefore choose a lengthscale $\ell^\star$ to nondimensionalize displacements by, but a different lengthscale $\sqrt{R^\star\ell^\star}$ to nondimensionalize distances by.  We therefore set
\begin{equation}\begin{aligned}
 \vect{u}^{\star} &= \ell^\star \vect{u},&
 \vect{v}^{\star} &= V^\star\vect{v}, &
 \eta^\star &= \ell^\star\eta, &
 \mu^\star &= \lambda^\star \mu, &
 \vect{x}^{\star} &= \sqrt{R^\star\ell^\star}\vect{x},
 \\
 t^{\star} &= \frac{\sqrt{R^\star\ell^\star}}{V^\star}t, &
 \mat{\tau^\star} &= \lambda^\star\sqrt{\frac{\ell^\star}{R^\star}}\mat{\tau}, &
 \rho^\star &= \frac{\lambda^\star}{{V^\star}^2}\rho &
 \epsilon &= \sqrt{\frac{\ell^\star}{R^\star}}, &
\end{aligned}\end{equation}
where $\lambda^\star$ and $\mu^\star$ are the Lam\'e coefficients for the elastic material, and $\epsilon$ will turn out to be the small parameter of interest.  Note that $\mu$ should not be confused with the coefficient of friction $\mu_0$.

\subsection{Linearization of the governing equations}\label{app:linearisation}

We now proceed to use the nondimensionalization above for the governing equations, under the assumption that the dimensionless parameter $\epsilon = \sqrt{\ell^\star\!/R^\star} \ll 1$.
The strain tensor $\mat{\varepsilon}$ becomes
\begin{equation}
\mat{\varepsilon} = \frac{1}{2}\big(\vect{\nabla^\star u^{\star}} + {\vect{\nabla^\star u^{\star}}}^T - \vect{{\nabla^\star u^\star}^T\cdot\nabla^\star u^\star}\big) = \epsilon\begin{pmatrix}
\frac{\partial u}{\partial x} & \frac{1}{2}\!\left(\frac{\partial u}{\partial y} + \frac{\partial v}{\partial x}\right)\\[1ex]
\frac{1}{2}\!\left(\frac{\partial u}{\partial y} + \frac{\partial v}{\partial x}\right) & \frac{\partial v}{\partial y}
\end{pmatrix} + O\big(\epsilon^2\big),
\end{equation}
and hence strains are small and we may assume linear elasticity.  Using the isotropic linear stress--strain relationship, the Cauchy stress tensor becomes
\begin{equation}
\mat{\tau} = \begin{pmatrix}
(1+2\mu)\frac{\partial u}{\partial x} + \frac{\partial v}{\partial y} & \mu\!\left(\frac{\partial u}{\partial y} + \frac{\partial v}{\partial x}\right) & 0\\[1ex]
\mu\!\left(\frac{\partial u}{\partial y} + \frac{\partial v}{\partial x}\right) & \frac{\partial u}{\partial x} + (1+2\mu)\frac{\partial v}{\partial y} & 0\\[1ex]
0 & 0 & \left(\frac{\partial u}{\partial x} + \frac{\partial v}{\partial y}\right)
\end{pmatrix}.
\end{equation}
The velocity equation~\eqref{equ:eulerian-velocity} becomes
 \begin{equation}
 \vect{v} = - \vect{e_x} + \epsilon\!\left(\frac{\partial \vect{u}}{\partial t} + \vect{v\cdot}\frac{\partial \vect{u}}{\partial \vect{x}}\right)
 = - \vect{e_x} + \epsilon\!\left(\frac{\partial \vect{u}}{\partial t} - \frac{\partial \vect{u}}{\partial x}\right)
 + O\big(\epsilon^2\big),
 \end{equation}
and hence the momentum equation~\eqref{equ:eulerian-momentum} simplifies to give
\begin{equation}
\rho\!\left(\frac{\partial}{\partial t} -\frac{\partial}{\partial x}\right)^{\!\!2}\!\!\vect{u}
= \vect{\nabla\cdot}\mat{\tau} + O\big(\epsilon^2\big).\label{equ:linearized-momentum}
\end{equation}
Hence, to leading order, the governing equations are the usual equations of linear elasticity, only with an extra advection term included:
\begin{subequations}\begin{align}
\rho\!\left(\frac{\partial}{\partial t} -\frac{\partial}{\partial x}\right)^{\!\!2}\!\!u
&= (1+2\mu)\frac{\partial^2 u}{\partial {x}^2} + (1+\mu)\frac{\partial^2 v}{\partial x\partial y} + \mu\frac{\partial^2 u}{\partial {y}^2},\\
\rho\!\left(\frac{\partial}{\partial t} -\frac{\partial}{\partial x}\right)^{\!\!2}\!\!v
&= (1+2\mu)\frac{\partial^2 v}{\partial {y}^2} + (1+\mu)\frac{\partial^2 u}{\partial x\partial y} + \mu\frac{\partial^2 v}{\partial {x}^2}.
\end{align}\label{app:equ:linearized-equations}\end{subequations}
It remains to apply the boundary conditions.

\subsection{Linearization of the boundary conditions}

Using the scaling above, the contact boundary condition~\eqref{equ:lagrangian-contact} with $y^{\star} = Y^{\star} + v^{\star}$ may be linearized directly.  Noting that $Y^{\star}=0$ on the surface, this leads to
\begin{align}
\big(\sqrt{R^\star\ell^\star} x \big)^2 + \big(R^\star - \ell^\star(v+1)\big)^2 & = R^2 &
&\Rightarrow &
v &= \tfrac{1}{2}{x}^2 - 1 + O\big(\epsilon^2\big),
\label{app:equ:linearized-contact}\end{align}
to be applied for $-a< x< d$.

The vertical displacement of the free surface $y^{\star} = \eta^{\star}(x^{\star})$ nondimensionalizes to give $y = \epsilon \eta(x) \ll 1$.  Expanding the slipping~\eqref{eqn: FSnonlinslip} and free~\eqref{eqn: FSfree} boundary conditions as a Taylor's series about $y = 0$ then gives
\begin{subequations}\begin{align}
\tau_{xy} \mp\mu_{0}\tau_{y y} &= 0 + O(\epsilon) && \text{for}\qquad y= 0, \qquad -a < x < d, \label{app:equ:linearized-friction}\\
\tau_{xy} = \tau_{yy} &= 0 + O(\epsilon) && \text{for}\qquad y= 0, \qquad x < -a \text{ and } x > d.\label{app:equ:linearized-free}  
\end{align}\end{subequations}
Note that the rescaled velocity of the roller is $v_c= R^\star\Omega^\star\!/V^\star$, and that the rescaled velocity of the elastic material in the same direction is $v_e= 1 + O(\epsilon)$, and so in the friction boundary condition~\eqref{equ:linearized-friction} we have $\pm = \mathrm{sgn}(R^\star\Omega^\star\!/V^\star-1)$.  The analysis above is invalid in the case $V^\star = R^\star\Omega^\star$; in this case the cylinder is no longer in the full slip case, and a more intricate mathematical model would be needed.

\subsection{Summary}

The linearized differential equations to solve are therefore~\eqref{app:equ:linearized-equations}, to be solved for $-\infty < x< \infty$ and $y< 0$, subject to boundary conditions along $y= 0$.  These boundary conditions are given by~\eqref{app:equ:linearized-contact} and~\eqref{app:equ:linearized-friction} for $-a< x< d$ when the elastic medium is in contact with the cylinder, and by~\eqref{app:equ:linearized-free} otherwise.

\bibliography{bib.bib}

\end{document}